\documentclass[]{aastex631}
\usepackage[normalem]{ulem}

\usepackage{CJKutf8}
\AtBeginDvi{\input{zhwinfonts}} 
\newcommand{\Chi}[2]{%
  \csname CJK*\endcsname{UTF8}{zhsong}%
    \CJKchar{#1}{#2}%
  \csname endCJK*\endcsname
}
\DeclareUnicodeCharacter{9673}{\Chi{"96}{"73}}
\DeclareUnicodeCharacter{82F1}{\Chi{"82}{"F1}}
\DeclareUnicodeCharacter{540C}{\Chi{"54}{"0C}}

\shorttitle{OSSOS Distant Resonant TNOs}
\shortauthors{Crompvoets et al.}

\begin{document}

\title{OSSOS XXV: Large Populations and Scattering-Sticking in the Distant Transneptunian Resonances}

\correspondingauthor{B. L. Crompvoets}
\email{bcrompvoets@uvic.ca}

\author[0000-0001-8900-5550]{B. L. Crompvoets}
\affiliation{Campion College and the Department of Physics, University of Regina, Regina, SK S4S 0A2, Canada}
\affiliation{Department of Physics and Astronomy, University of Victoria, Victoria, BC, Canada}
\author[0000-0001-5368-386X]{S.~M. Lawler}
\affiliation{Campion College and the Department of Physics, University of Regina, Regina, SK S4S 0A2, Canada}
\author[0000-0001-8736-236X]{K.~Volk}
\affiliation{Lunar and Planetary Laboratory, University of Arizona, 1629 E University Blvd, Tucson, AZ 85721, USA}
\author[0000-0001-7244-6069]{Y.-T. Chen
(\Chi{"96}{"73}\Chi{"82}{"F1}\Chi{"54}{"0C})}
\affiliation{Institute of Astronomy and Astrophysics, Academia Sinica; 11F of AS/NTU Astronomy-Mathematics Building, No. 1 Roosevelt Rd., Sec. 4, Taipei 10617, Taiwan}
\author[0000-0002-0283-2260]{B.~Gladman}
\affiliation{Department of Physics and Astronomy, University of British Columbia, Vancouver, BC V6T 1Z1, Canada}
\author[0000-0002-9179-8323]{L.~Peltier}
\affiliation{Campion College and the Department of Physics, University of Regina,
Regina, SK S4S 0A2, Canada}
\affiliation{Department of Physics and Astronomy, University of Victoria, Victoria, BC, Canada}
\author[0000-0003-4143-8589]{M.~Alexandersen}
\affiliation{Center for Astrophysics $|$ Harvard \& Smithsonian, 60 Garden Street, Cambridge, MA 02138, USA}
\author[0000-0003-3257-4490]{M.~T. Bannister}
\affiliation{School of Physical and Chemical Sciences | Te Kura Mat\={u}, University of Canterbury,
Private Bag 4800, Christchurch 8140, New Zealand}
\author[0000-0001-8221-8406]{S.~Gwyn}
\affiliation{Herzberg Astronomy and Astrophysics Research Centre, National Research Council of Canada, 5071 West Saanich Rd, Victoria, BC V9E 2E7, Canada}
\author[0000-0001-7032-5255]{J.~J. Kavelaars}
\affiliation{Herzberg Astronomy and Astrophysics Research Centre, National Research Council of Canada, 5071 West Saanich Rd, Victoria, BC V9E 2E7, Canada}
\affiliation{Department of Physics and Astronomy, University of Victoria, Victoria, BC, Canada}
\author[0000-0003-0407-2266]{J.-M.~Petit}
\affiliation{Institut UTINAM UMR6213, CNRS, Univ. Bourgogne Franche-Comt\'e, OSU Theta F25000 Besan\c{c}on, France}

\begin{abstract}

There have been 77 TNOs discovered to be librating in the distant transneptunian resonances (beyond the 2:1 resonance, at semimajor axes greater than 47.7~AU) in four well-characterized surveys: the Outer Solar System Origins Survey (OSSOS) and three similar prior surveys. Here we use the OSSOS Survey Simulator to measure their intrinsic orbital distributions using an empirical parameterized model.
Because many of the resonances had only one or very few detections, $j$:$k$ resonant objects were grouped by $k$ in order to have a better basis for comparison between models and reality.
We also use the Survey Simulator to constrain their absolute populations, finding that they are much larger than predicted by any published Neptune migration model to date; we also find population ratios that are inconsistent with published models, presenting a challenge for future Kuiper Belt emplacement models. 
The estimated population ratios between these resonances are largely consistent with scattering-sticking predictions, though further discoveries of resonant TNOs with high-precision orbits will be needed to determine whether scattering-sticking can explain the entire distant resonant population or not.

\end{abstract}


\section{Introduction}
\pagenumbering{arabic}

Observations of the outer solar system indicate an enormous number of objects trapped within the very distant mean-motion resonances with Neptune. This work aims to provide a greater understanding of the absolute populations in these different resonances, the distribution of orbital elements within each resonance, as well as to help identify the primary mechanism by which these distant resonances are populated. Population of these distant resonances may occur through sweep-up during a smooth phase of Neptune's migration \citep[e.g.][]{Hahn:2005}, during circularization of Neptune's orbit after a Nice Model-style instability \citep[e.g.][]{Levison:2008}, or may be primarily populated by ``scattering-sticking'': unstable scattering TNOs temporarily sticking within resonances \citep[e.g.][]{Yu:2018}.

Previous papers \citep[e.g.,][]{gladman2012resonant,Pike:2015,Alexandersen:2016,Volk:2016,Volk:2018,Chen:2019} have already modelled many of the resonant populations in detail, however, all of the distant high-order resonant populations have not yet been modelled. This paper shall refer to all resonances beyond the 2:1 resonance at 47.7~AU as `distant' resonances. 

Several of the distant resonances between 2:1 and 5:1 were modelled by \citet{gladman2012resonant}, who measured large population estimates. 
Individual analyses of the 3:1 \citep{Alexandersen:2016}, 4:1 \citep{Lawler2013,Alexandersen:2016}, 5:1 \citep{Pike:2015}, and 9:1 \citep{Volk:2018} have also been carried out. 
However, these are only a few of the many resonances for which we have detections and thus modelling capability. 

Previous theoretical works suggest mechanisms which could have built up the resonant populations within the distant Solar System. \citet{KaibSheppard2016} state that resonant objects beyond the 4:1 resonance are less strongly affected by Neptune's migration due to Kozai cycling timescales. Furthermore, \citet{Pike:2015, Volk:2018} both found that the large libration amplitudes of the small number of discovered 5:1 and 9:1 resonators were consistent with scattering-sticking, but the absolute population implied by the observed TNOs was larger than could be easily explained by scattering-sticking. 
These previous results point to distant resonances being populated by TNOs from the scattering population sticking to resonances rather than being swept up during Neptune's migration itself, although whether the relative population sizes within resonances supports this theory remains unclear.

\citet{Yu:2018} extensively modelled the temporary sticking of scattering TNOs within the region $a$=30-100 AU, with orbital distributions of initial populations based off the \citet{Kaib:2011} scattering model and constraints on the current total population of scattering objects \citep{Lawler2018scattering}. 
After measuring the time-averaged population over 1 Gyr, they found that, at any time, approximately 40\% of all objects within this semi-major axis range should be stuck in a resonance. Their simulations imply that a) approximately 1.3 times as many non-resonant scattering TNOs exist as those transiently stuck, b) a quarter of those that are stuck should exist in the $n:1$ resonances, and c) that the resonant-sticking populations should increase with semi-major axis \citep[well demonstrated in][]{Volk:2018}. 

In this work, we create empirical parametric models which reproduce the observed orbits of the distant, high-order resonances in the transneptunian belt. The properties of objects within these resonances have generally not previously been measured due to the extreme difficulty of observing these distant bodies, measuring their orbits to high enough precision for resonant identification, and accounting for the complex observation biases. 
The rate and mode by which Neptune migrated outwards, combined with ongoing resonant sticking by the scattering population \citep[see, e.g.,][]{LykawkaMukai2007,Yu:2018}, affects the distribution of objects among the different resonances that we observe today, although the relative importance of each type of emplacement, particularly for the distant resonances, is as yet unmeasured. N-body simulations show that the distribution of TNOs in resonances is different depending on which type of migration Neptune experiences \citep[e.g.][]{Hahn:2005,Levison:2008,Nesvorny:2016,KaibSheppard2016}. When the final test particles are dynamically classified \citep[as in e.g.][]{Pikeetal2017,Lawler:2019}, these models can be tested by comparison to observations from well-calibrated surveys \citep[as in e.g.][]{gladman2012resonant}. 

We make use of precise orbital measurements of transneptunian objects (TNOs) from the Outer Solar System Origins Survey \citep[OSSOS;][]{Bannister:2018} in combination with three other well-characterized surveys: the Canada France Ecliptic Plane Survey \cite[CFEPS;][]{Petit:2011}, the CFEPS High Latitude Survey \citep[HiLat;][]{Petit:2017} and the \citet{Alexandersen:2016} survey. These four surveys together will be referred to as OSSOS++. We call these surveys ``well-characterized'' because all biases in pointing direction, magnitude limits, detection, and tracking are well-known and can be incorporated into a Survey Simulator\footnote{The OSSOS Survey Simulator is publicly available at \url{https://github.com/OSSOS/SurveySimulator}.} \citep{Lawler2018survey,Petit:2018}. We use the Survey Simulator to subject orbital models to the same biases as the observations, allowing for robust comparisons between simulated and real detections to test the models.

In this paper, we produce a set of parameterized orbital distributions which are statistically consistent with our well-characterized distant resonant TNO observations (Sections~\ref{sec:orbits} and \ref{sec:results}). We then compare our measured populations with previously published N-body simulations of the Kuiper Belt, finding that none reproduce the quantity calculated in the following population models (Section~\ref{sec:disc}). 
Future Solar System dynamical models must match our population models constrained by observations within their uncertainty, and this will help determine the most likely mechanism which emplaced TNOs into the distant Neptune resonances.

\section{Measuring Orbital Distributions and Populations} \label{sec:orbits}

\begin{figure}
\centering
\includegraphics[scale=0.6]{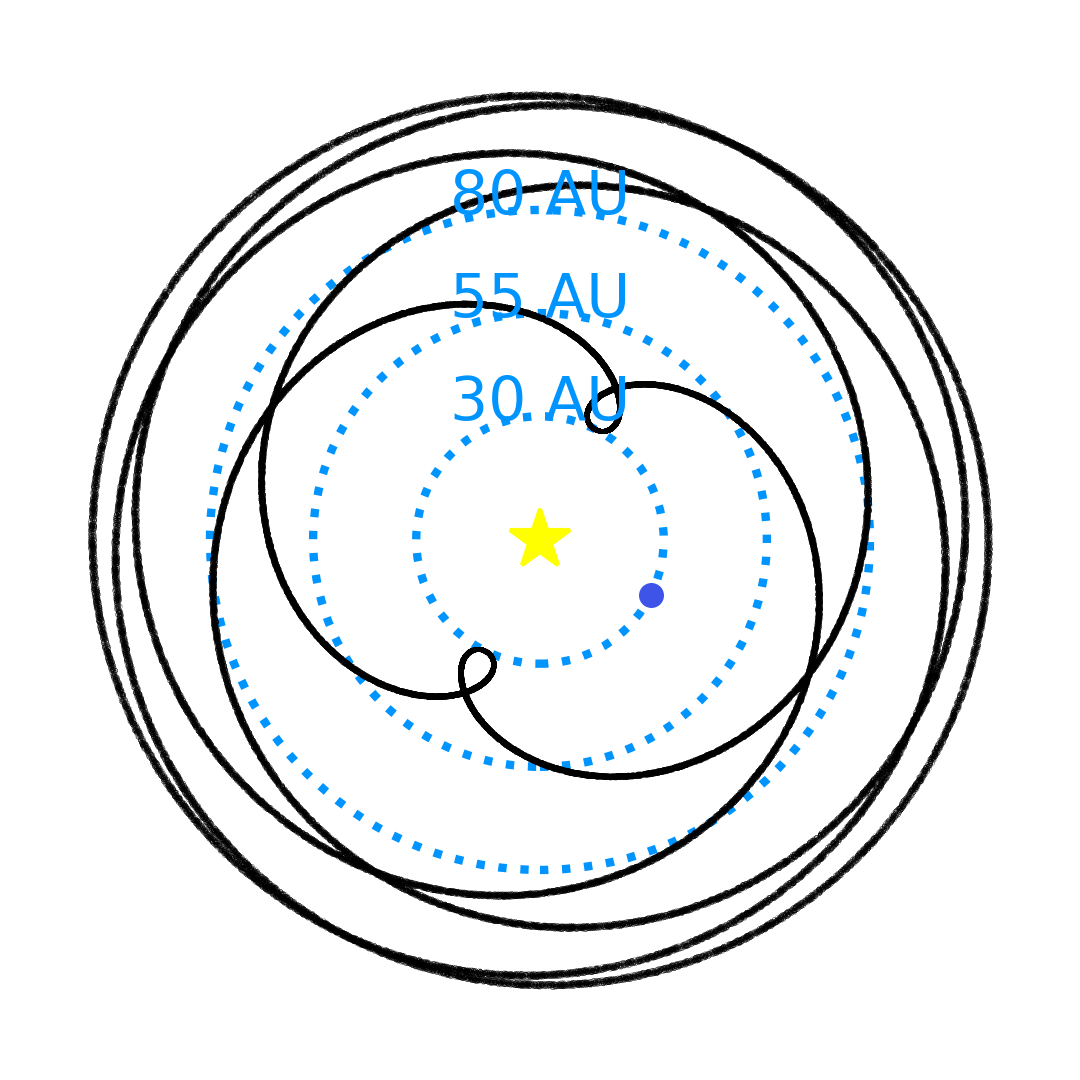}\includegraphics[scale=0.6]{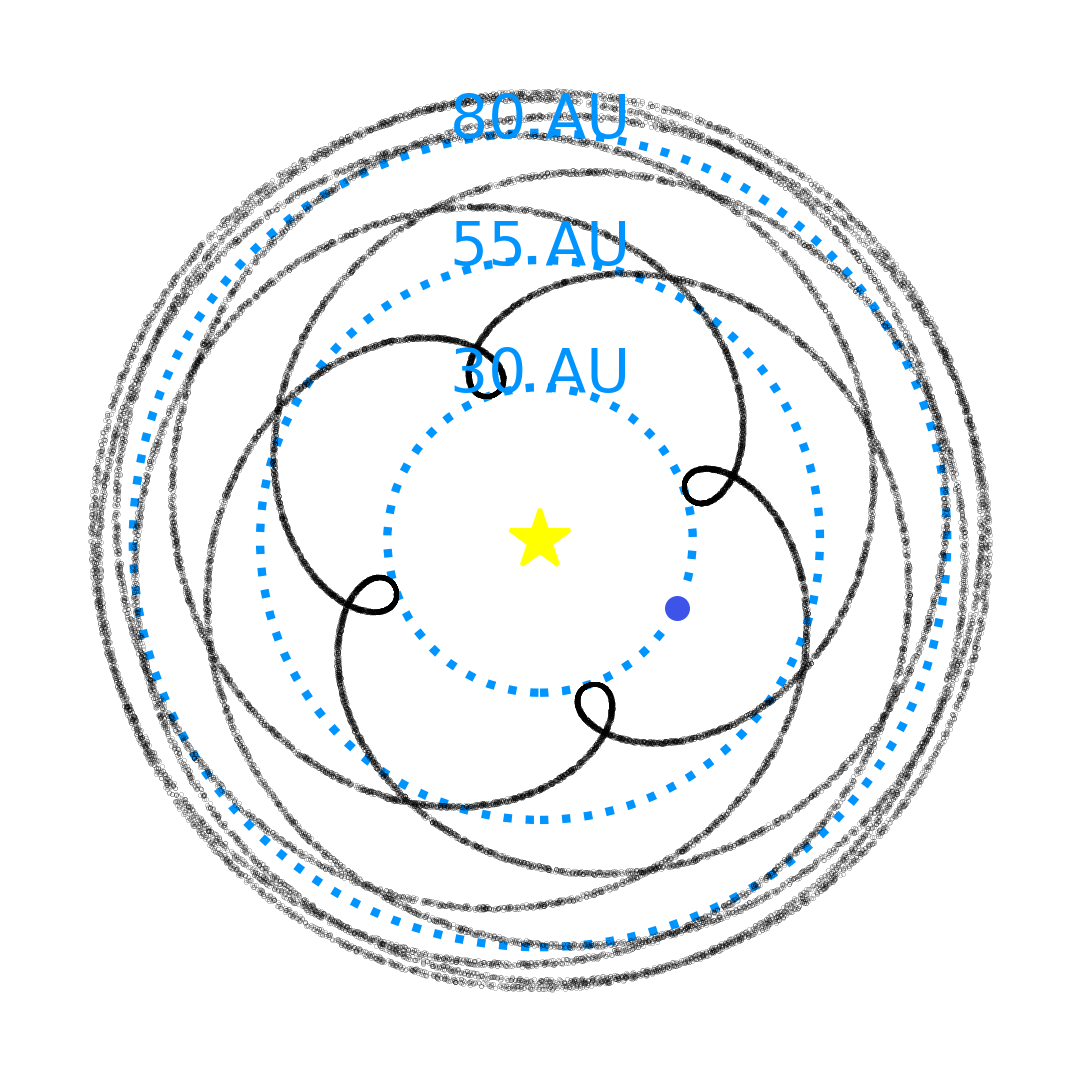}\includegraphics[scale=0.6]{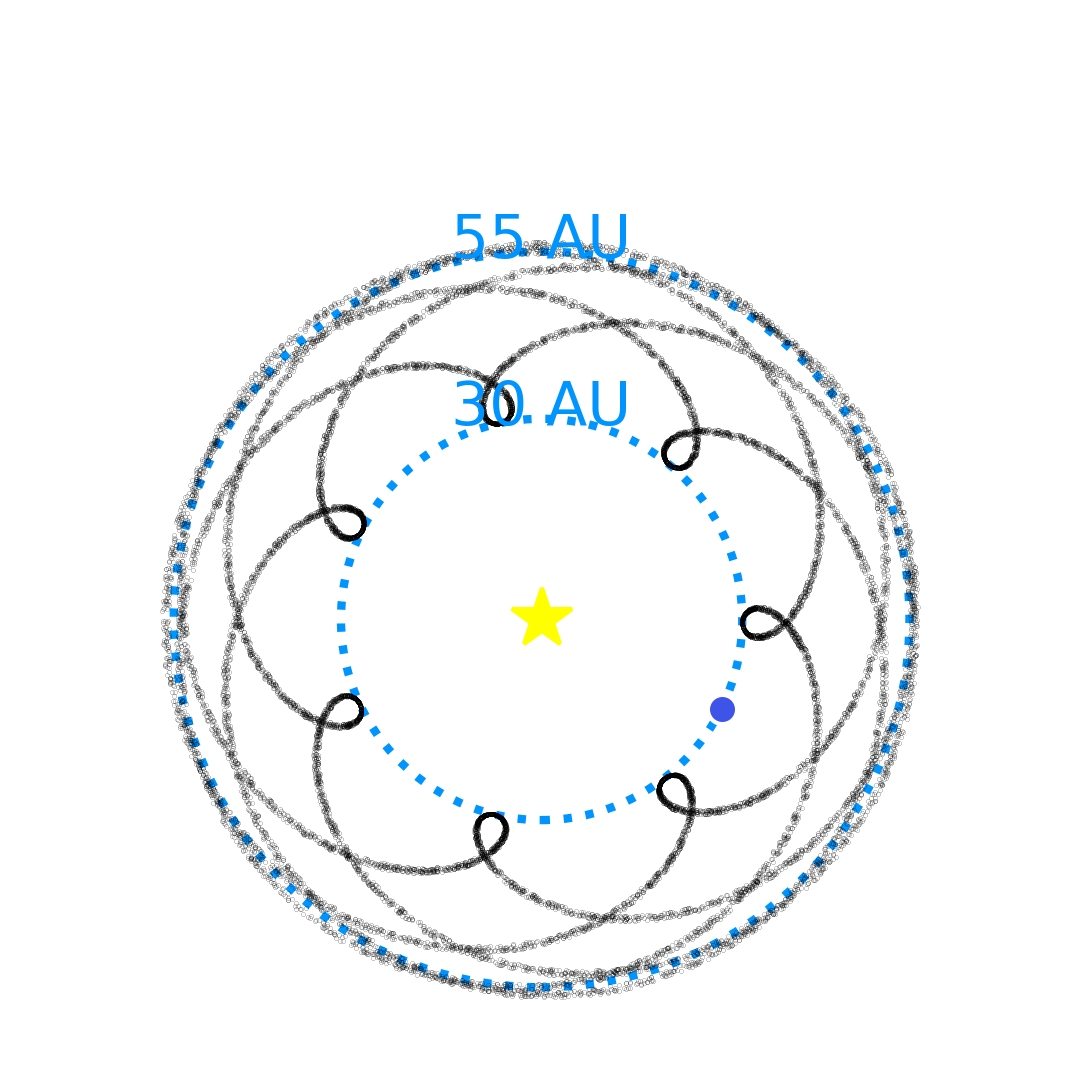}
\includegraphics[scale=0.6]{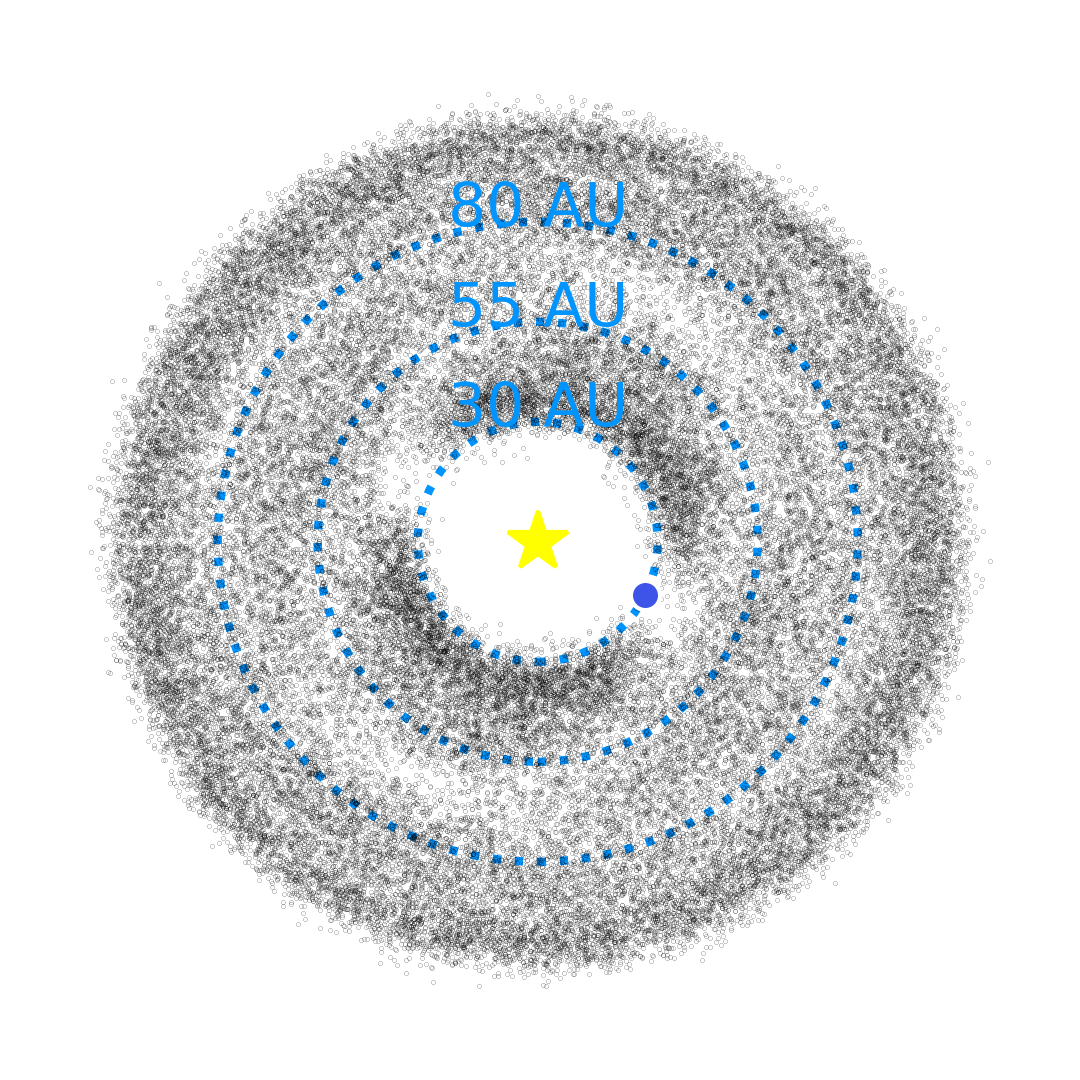}\includegraphics[scale=0.6]{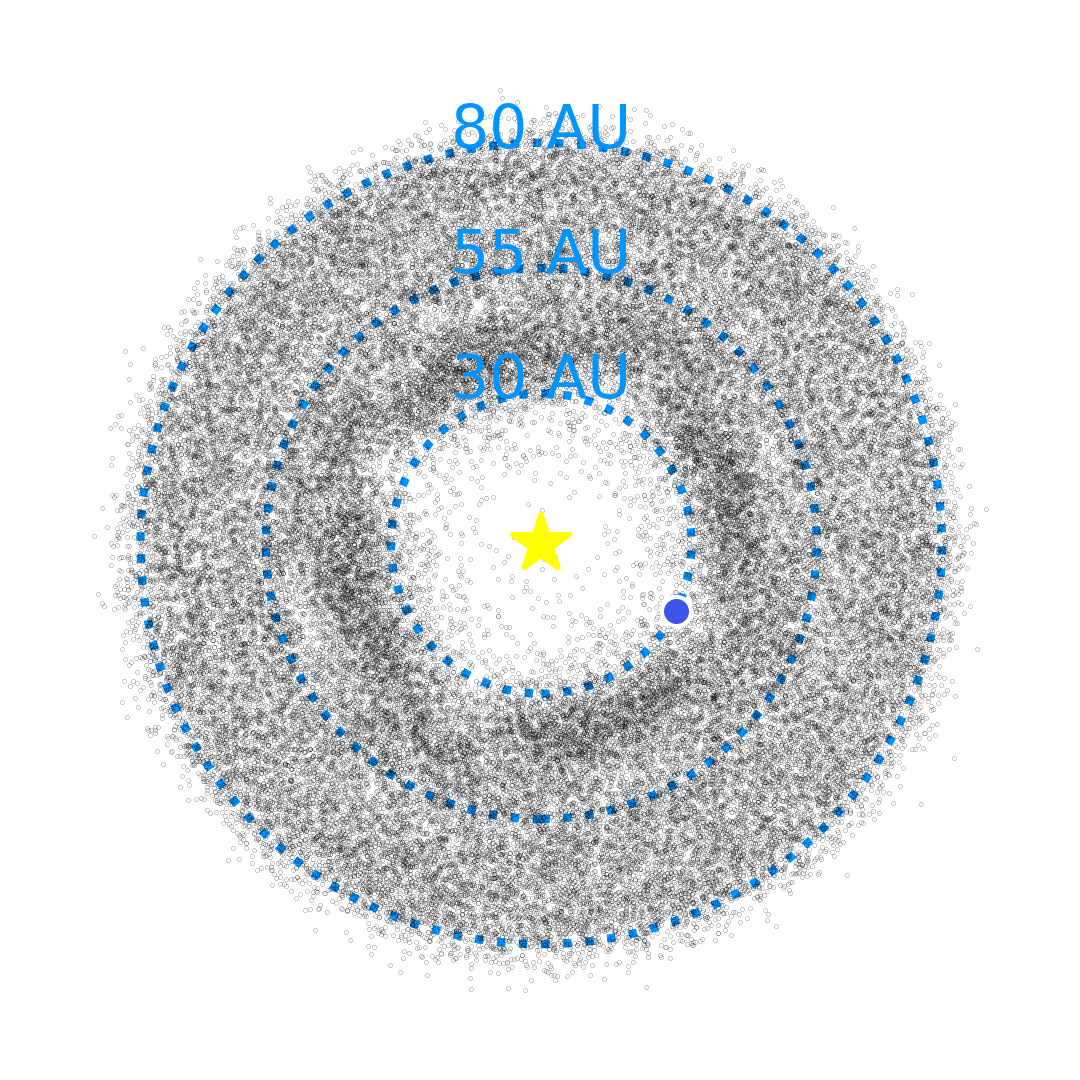}\includegraphics[scale=0.6]{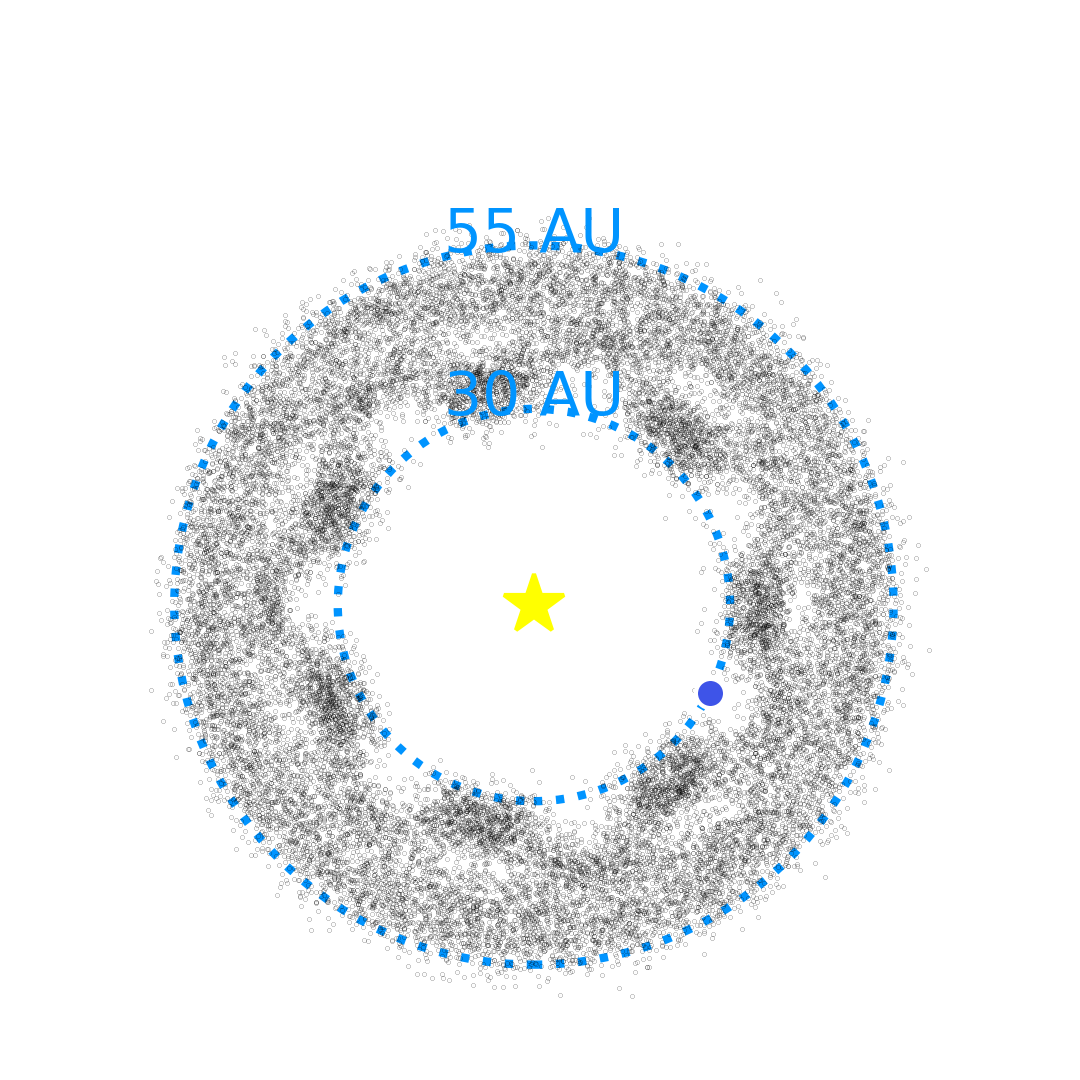}
\caption{Top 3 panels: Toy models of 7:2 ($a=69~$AU), 11:4 ($a=59$~AU), and 12:7 ($a=43$~AU) resonances demonstrate the equilibrium configurations of the resonances with respect to Neptune (blue point).
Bottom 3 panels: the 7:2, 11:4, and 12:7 resonances with realistic orbital distributions, where libration amplitude can be large and thus the pericenter location oscillates about the equilibrium configuration. Depending on which resonance one looks at, its pericenter location (and thus the most likely location to detect TNOs) differs. For these toy models, pericenter occurs at Neptune's semi-major axis.}

\label{fig:resj:k}
\end{figure}

An object is said to be in a mean-motion resonance with Neptune when its orbital period around the Sun can be written as a ratio of relatively small integers relative to Neptune's orbital period and the repeating perturbations on the object's orbit by Neptune causes the object to librate around the exact resonant orbit. 
For example, Pluto is in the 3:2 resonance with Neptune; Pluto goes around the Sun twice in the same amount of time it takes Neptune to go around the Sun three times, and the two perihelion passages Pluto makes during this resonant cycle librate around points $90^\circ$ to either side of Neptune.
This is a simplified description of the complex phenomena that can occur in Neptune's resonances (which include higher-order sub-resonances within these mean-motion resonances), but in general, an object in an exact $j$:$k$ external resonance with Neptune will come to perihelion at $k$ different locations relative to Neptune during each resonance cycle, as illustrated in the top panels of Figure~\ref{fig:resj:k}. 
In reality, a resonant object will librate around these $k$ points, somewhat smearing out these perihelion locations in a population of resonant objects (bottom panels of Figure~\ref{fig:resj:k}).
TNOs on unstable, scattering orbits can ``stick'' to resonances temporarily; objects experiencing even quite distant encounters with Neptune can experience changes in semimajor axis that place them into temporarily resonant configurations with the planet. Neptune can then provide the necessary gravitational nudges to hold the TNO in the resonant orbit for anywhere from thousands to hundreds of millions of years before the TNO finally leaves the resonance and returns to an unstable scattering orbit \citep[see][for more a detailed discussion]{Yu:2018}.

For an integer orbital period ratio $j$:$k$, we identify resonant behavior by examining the resonant angle 
\begin{equation}
    \phi_{jk}=j\lambda_{\rm TNO}-k\lambda_{\rm N}-(j-k)\varpi_{\rm TNO}, \label{eq:phi}
\end{equation} 
which for resonant objects must be confined and thus not take on all values between $0^{\circ}-360^{\circ}$; this confinement is directly related to the libration around the equilibrium points described above.
In Equation~\ref{eq:phi}, the mean longitude $\lambda=\Omega+\omega+\mathcal{M}$ is the combination of the orbital angles of the longitude of ascending node $\Omega$, the argument of pericenter $\omega$, and the mean anomaly $\mathcal{M}$; the longitude of pericenter $\varpi=\Omega+\omega$, and the subscripts N and TNO refer to Neptune and the TNO respectively.
In principle, resonant arguments involving combinations of $\varpi_{\rm TNO}$ and $\Omega_{\rm TNO}$ are possible for each $j$:$k$ resonance. 
However, these distant resonant TNOs typically have high eccentricities and so the combination of arguments in equation~\ref{eq:phi} is strongest \citep[e.g.,][]{Murray}, and we use this equation for our resonant modelling.

This confinement of orbital angles has a powerful effect on the observability of resonant TNOs:
the confinement of $\phi_{jk}$ means that resonant TNOs will have their pericenters occur preferentially at set locations on the sky relative to Neptune \citep[for a detailed example, see][]{Lawler:2013}.
In the frame of reference that is co-rotating with Neptune, the orbits of resonant objects look like a child's spirograph, mainly circular but with small loops at pericenter (top portion of Figure \ref{fig:resj:k} shows toy models of the 7:2, 11:4, and 12:7 resonances that demonstrate this). 
Because surveys are most sensitive to TNOs at perihelion 
(as they are then closest to the Sun and thus brightest)
, and different resonances have concentrations of perihelia at different sky locations, the sensitivity of a given survey to resonant TNOs can only be evaluated by including the on-sky pointings and completeness functions of the survey in careful modelling.

\subsection{OSSOS++ TNO Orbits}
 
OSSOS was specifically designed to detect and track many resonant TNOs to high orbital precision, and was successful in this goal.  Orbital resonance has been diagnosed by integrating an observed TNO's best-fit orbit as well as two clones of that orbit that represent the maximum and minimum semi-major axes consistent with the observations, for 10~Myr under the gravitational influence of the Sun and the four giant planets.  
We then inspect for libration of the resonant angle for a very large set of possible Neptune resonances. This is done by
visually inspecting the time histories of all potentially librating angles identified by a simple automated algorithm. Next, we visually examine the semimajor axis evolution of all observed objects for signs of resonant behavior not identified by the automated algorithm 
\citep[following the classification scheme of][]{Gladman2008}.  If all three clones of a TNO librate in the resonance for more than half of the 10~Myr integration, 
the TNO is securely resonant (in practice, most of the resonant objects librate cleanly the entire 10 Myr, though a minority of resonant-classified objects experience relatively stable, though intermittent, libration).  If at least one clone does not meet that libration criteria, but the best-fit orbit does,  then it is designated as ``insecurely'' resonant. All TNOs along with their label as either secure or insecure are given in Appendix~\ref{sec:tnolist}.
 Future observations of the insecurely resonant TNOs will help to increase their orbital precision and diagnose whether they are currently resonant or not. 

In this work, we consider 48 of the 77 discovered distant resonant TNOs from OSSOS++ (Table~\ref{tab:listofres}). This is all of the resonant TNOs that are at larger semimajor axes than the 2:1 ($a\simeq47.7$~AU, see \citet{Chen:2019} for modelling of the 2:1), but excluding those in the 5:2, which was thoroughly modelled in \citet{matthews_2019} (we include their population measurements alongside ours for completeness).  Of these 48 resonant TNOs, 25 are securely resonant and 23 are insecurely resonant, belonging to 22 separate resonances. A full list of MPC designations, orbital elements, $H$ magnitudes, and classifications for these resonant TNOs is given in Appendix~\ref{sec:tnolist}.   
Full details on the observing strategy and dynamical classification of all OSSOS TNOs are discussed in \citet{Volk:2016} and \citet{Bannister:2018}, and we have applied those same criteria to the dynamical classifications of the other OSSOS++ TNOs included in this work.

\begin{deluxetable}{ccccccc}												
\tablecaption{Summary of OSSOS detections.\label{tab:listofres}}												
\tablewidth{0pt}												
\tablehead{												
\colhead{n:1 (22)}	&	\colhead{n:2 (33)}	&	\colhead{n:3 (6)}	&	\colhead{n:4 (5)}	&	\colhead{n:5 (6)}	&	\colhead{n:6 (3)}	&	\colhead{n:8 (2)}
}												
\startdata												
3:1 (12)	&	5:2 (29)	&	7:3 (3)	&	11:4 (2)	&	11:5 (2)	&	13:6 (2)	&	17:8 (1)\\
4:1 (5)	&	7:2 (2)	&	8:3 (2)	&	15:4 (1)	&	12:5 (2)	&	23:6 (1)	&	35:8 (1)\\
5:1 (3)	&	9:2 (1)	&	10:3 (1)	&	17:4 (1)	&	13:5 (1)	&		&	\\
9:1 (2)	&	23:2 (1)	&		&	27:4 (1)	&	24:5 (1)	&		&	
\enddata												
\tablenotetext{}{These are the resonances with OSSOS++ detections that are more distant than the 2:1, organized by $k$, with the number of OSSOS++ detections for each $j$:$k$ resonance given in parentheses. Note that the 5:2 has many OSSOS++ detections - these are modelled in detail by \citet{matthews_2019}, and we do not repeat that analysis.}												
\end{deluxetable}												
 
\subsection{The Survey Simulator}
 
In order to constrain the true populations and orbital distributions of TNOs within the different mean-motion resonances represented in the OSSOS++ observed TNO sample, the Survey Simulator is necessary to impose the survey's observational biases on model populations and allow statistical comparison. The Survey Simulator is a piece of software  
built by the OSSOS team \citep{Petit:2018,Lawler2018survey} which applies the OSSOS++ survey biases to a model population. This allows direct comparison between the orbital properties of real discovered TNOs and simulated detections from a survey-biased model. These comparisons allow us to search for a parametric orbital model that is statistically consistent with each observed TNO population.
 
We use the Survey Simulator to test our parameterized orbital models to ensure that they are statistically consistent with the OSSOS++ resonant TNO detections. 
The Survey Simulator generates simulated TNOs from a given model continuously until a predetermined number of objects have been `detected' by the characterization of the OSSOS++ surveys, i.e. the modelled object exists within the phase space of on-sky position, observation time, TNO brightness, and rate of motion on the sky which the OSSOS++ surveys would have been able to detect. 
Once a parameterized orbital distribution is obtained which accurately reproduces the observed data, it can be compared with the outcomes from different planetary migration models.
These parameterized models can also be used to measure the absolute populations in different resonances. 

\subsection{Parameterized Orbital Models} \label{sec:parameterized}

Similar to previous works \citep[e.g.,][]{gladman2012resonant,Volk:2016}, we empirically parameterize each orbital element distribution within a resonance independently - we know that this is not fully accurate, as some elements are likely to be correlated within resonant populations \citep[e.g.][]{Morbidelli1997,Tiscareno2009}. But for the small number of detections in each resonance, this parameterization will be more than sufficient to produce a statistically acceptable model. The simple empirical parameterizations we use here are based on parameterizations of resonances with more known members \citep[e.g.,][]{gladman2012resonant,Alexandersen:2016,Volk:2016}, and have been used successfully to model resonances with only a few known members \citep[e.g.,][]{Bannister:2016,Volk:2018}. We note that these orbital parameterizations are primarily based on the closer-in resonances, and if TNOs are primarily emplaced in the distant resonances by a different pathway, they may have different orbital distributions than the inner resonances. However, due to the small number of know objects in the resonances we are modelling here, we build on previous work rather than independently fitting new parameterizations. We use the following empirical parameterizations:

The semi-major axis is chosen to be uniformly random within $\pm$0.5~AU of the resonance center, where the resonance center is the precise semi-major axis  to create the necessary orbital period ratio with Neptune. For the $j$:$k$ resonance, this is:
\begin{equation}
    a_{jk}=a_{\rm N}\left(\frac{j}{k}\right)^{2/3}
\end{equation}
where $a_N$ is the average semimajor axis of Neptune.  We note that this neglects the eccentricity-dependent width of the resonances, but such detailed modeling of the a distribution is not important for detectability within the resonance.

The eccentricity is determined via a parameterization of the pericenter $q$ distribution, related via $q=a(1-e)$. Based on previous orbital distribution modelling \citep[e.g.,][]{gladman2012resonant} which successfully used a Gaussian distribution in $e$, with a semimajor axis $a_{jk}$ as given above for each resonance. Here we model the pericenter distribution as a Gaussian centered on a given central value $q_c$, with width $q_w$:
\begin{equation}
    P(q)~ {\displaystyle \propto } ~\exp{\frac{-(q-q_{c})^2}{2\times(q_w)^2}}
    \label{eq:q}
\end{equation}

The inclination distribution is chosen to reflect the standard $\sin{i}$ times a Gaussian \citep{Brown2001} with a given inclination width $\sigma_i$:
\begin{equation}
    P(i) ~{\displaystyle \propto}~  \sin{i} \exp{\frac{-i^2}{2\times(\sigma_i)^2}}
    \label{eq:i}
\end{equation}

We note that in this parameterization, we have ignored any Kozai component in these distant resonant populations.  Kozai resonators have been diagnosed in high-order resonances sunward of the 2:1 \citep[e.g.,][]{LykawkaMukai2007,Lawler:2013}, but they make up a very small fraction of the TNOs within each resonance. 
The Kozai orbital phase space in the resonances is also difficult to model accurately using even relatively complex parameterizations (see discussion in \citealt{Volk:2016}), 
so we ignore the current Kozai component for now.  While the fraction of each resonant population that is currently experiencing Kozai oscillations is likely small, the distribution of orbital elements within each resonance has likely been affected by past Kozai cycling, as has been shown in detailed analyses of dynamical simulations \citep[e.g.][]{PikeLawler:2017,Lawler:2019}.  
Past Kozai cycling within a resonance should lead to an $e$-$i$ distribution that is on average anti-correlated, with low-$i$ resonant TNOs more likely to have high-$e$. This adds an additional potential observing bias since the low-$e$, low-$i$ TNOs will be most detectable in primarily ecliptic plane surveys like those that make up OSSOS++.  With so few detections, and poorly known phase space distributions for these distant resonances, we feel that attempting to model a correlation between the $e$ and $i$ distributions would be over-fitting, and hope to see this detailed modelling become viable with more detections in future surveys.

For all resonances except the $n$:1 resonances, the libration amplitudes are chosen from the best-fit probability distribution for the plutinos as determined in \citet{gladman2012resonant}, with the most-likely libration amplitude being 95$^{\circ}$, and a range of $20^\circ - 130^\circ$.  We did not fit for different libration amplitude distributions, but relied on previous measurements of more easily studied close-in resonances.
As in \citet{gladman2012resonant} and \citet{Chen:2019}, the resonant angle (equation~\ref{eq:phi}) is chosen differently for the $n$:1 resonances, where there is more than one possible libration center: each simulated object must first be assigned to the symmetric resonant island or the leading or trailing asymmetric resonant island, which have different resonance centers and libration amplitudes. The population fraction of $n$:1 resonators residing in the symmetric island is chosen to be 30\%, with the remaining 70\% being split equally into leading/trailing asymmetric islands. This is based on previous modelling \citep[e.g.][]{gladman2012resonant,Chen:2019}, and though the fraction of objects in each $n$:1 island is not currently well-constrained by observations, we do not expect this to significantly affect our population estimates.
We know from dynamical models of Neptune's $n$:1 resonances \citep[e.g.,][]{Nesvorny:2001}, confirmed from detailed modelling detections in the 2:1 \citep[e.g.][]{Chen:2019}, that the orbital element distributions are different in each $n$:1 resonant island (for example, the exact resonant center is dependent on eccentricity in the asymmetric islands).
However, given the limited number of detections we are working with in each of the $n$:1 resonances, this level of detail is not necessary, and we used simple parameterizations (we note, however, that more detailed modelling may become important as more resonant TNOs are discovered in the future). All other resonances have only a single libration center at $\phi_{jk}=180^{\circ}$.  

After the resonance center and libration amplitudes are selected, the resonant angle is chosen sinusoidally from within the libration amplitude, so that resonant angles closer to the libration amplitude limits are more likely to be chosen. This is because TNOs spend more time close to the extrema of their libration amplitudes, and a sinusoidal distribution reproduces this effect well \citep{Volk:2016}.
Next, the longitude of ascending node $\Omega$ is chosen randomly between $0^{\circ}$ and $360^{\circ}$. 
In order to have the resonant argument represent all possible mutual configurations, the mean anomaly $\mathcal{M}$ is chosen randomly between $0^{\circ}$ and ($k\times360^{\circ}$). Equation~\ref{eq:phi} is then used to calculate the argument of pericenter $\omega$ to satisfy the resonant condition for that particular $j$:$k$ resonance. 

To calculate the brightness of a simulated object, the distance is combined with the absolute $H$-magnitude (we use $H_r$, the absolute magnitude in $r$-band, to be consistent with other OSSOS++ works\footnote{Several of the TNOs in this sample were not observed in $r$-band because some blocks of CFEPS observed only in $g$.  These have had their $g$-band and $H_g$ magnitudes transposed to $r$ by assuming that $g-r=0.7$, which is at the neutral end of the observed color range of dynamically excited TNOs \citep{Tegleretal2016}. We note that \citet{Shankman:2016} used $g-r=0.7$, and also showed that using $g-r$ values ranging from 0.5--0.9 makes no difference to the statistical analysis \citep[see Figure~8 in][]{Shankman:2016}}.).  We assigned $H_r$ magnitudes to simulated objects by drawing from the best-fit divot size distribution from \citet{Lawler2018scattering}:
\begin{equation}
    \frac{dN}{dH} \propto 10^{\alpha H}
    \label{eq:H}
\end{equation}
with $\alpha=0.9$ for $H_r<8.3$ and $\alpha=0.5$ for $H_r>8.3$, with a contrast factor of 3.2 at the break value ($H_r=8.3$), as explained in \citet{Lawler2018scattering}.

\subsection{Constraining Models with Small Numbers of Detections} \label{sec:smallnum}

Very detailed parametric models of individual resonant populations can only be  achieved for resonances with large numbers of detections \citep[e.g.][]{Chen:2019,matthews_2019,Lin:2021}. Previous works on resonant populations with smaller numbers of detections have typically constrained more simplified parametric models \citep[e.g.][]{gladman2012resonant,Alexandersen:2016,Volk:2016} similar to those described in Section 2.3. In order to produce population estimates for resonances with one to a few detections, the eccentricity and inclination distributions have typically been assumed (based on dynamical considerations or other nearby resonances) rather than fit  \citep[e.g.][]{Pike:2015,Volk:2018}.
Because of the very small number of detections spread across many distant, high-order resonances (Table~\ref{tab:listofres}), much of the analysis in this work is done by assuming that groups of resonances have similar distributions, and fitting inclination and pericenter distributions in groups across multiple resonances. We adopt this method to avoid over-fitting our sparse data, while still calculating statistically acceptable orbital parameterizations. 
Scattering-sticking resonance modelling \citep[e.g.][]{LykawkaMukai2007,Yu:2018} reveals that for high-eccentricity TNOs, the value of $k$ in a $j$:$k$ resonance is more important than the order ($j-k$) of a resonance for determining its strength (see detailed discussion in \citealt{Volk:2018}), which is why we grouped resonances in this manner.
 
We follow the same basic procedure as for previous OSSOS++ analyses \citep[e.g.][]{KavelaarsL3,gladman2012resonant,Shankmanetal2013,Alexandersen:2016,Volk:2016,Lawler2018scattering}. The Anderson-Darling test \citep{Anderson-Darling} was used to statistically evaluate the validity of our parameterized orbital models by comparing the simulated detections to the real detections. This test is similar to the better-known Kolmogorov-Smirnov test statistic, but is more weighted toward the tails of the distribution, and provides information on whether or not a model is statistically rejectable. We run the parameterized orbital models through the Survey Simulator, creating a distribution of simulated detections. The Anderson-Darling statistic is then calculated between the simulated detections and the real OSSOS++ detection in each orbital model parameter (primarily focusing on $q$ and $i$, see Figure~\ref{fig:n2cumu} for an example).  Rather than rely on tables of critical values, we bootstrap the Anderson-Darling statistic by randomly drawing 100 samples from the simulated detections of the same size as the number of real TNOs. We then compare the AD statistic for these subsamples tested against the total simulated detection sample to the AD statistic for the real TNOs tested against the total simulated detections.  The fraction of random samples with a larger AD statistic than the real TNO sample is the bootstrapped Anderson-Darling (AD), or rejectability, statistic. Any model which provides a rejectability value $\geq5\%$ for all tested parameters is considered statistically consistent, regardless of whether the rejectability value is 6\% or 96\%. 
 
 \section{Results} \label{sec:results}

\subsection{A Parameterized Orbital Distribution}

\begin{figure}
    
    \includegraphics[width=\linewidth]{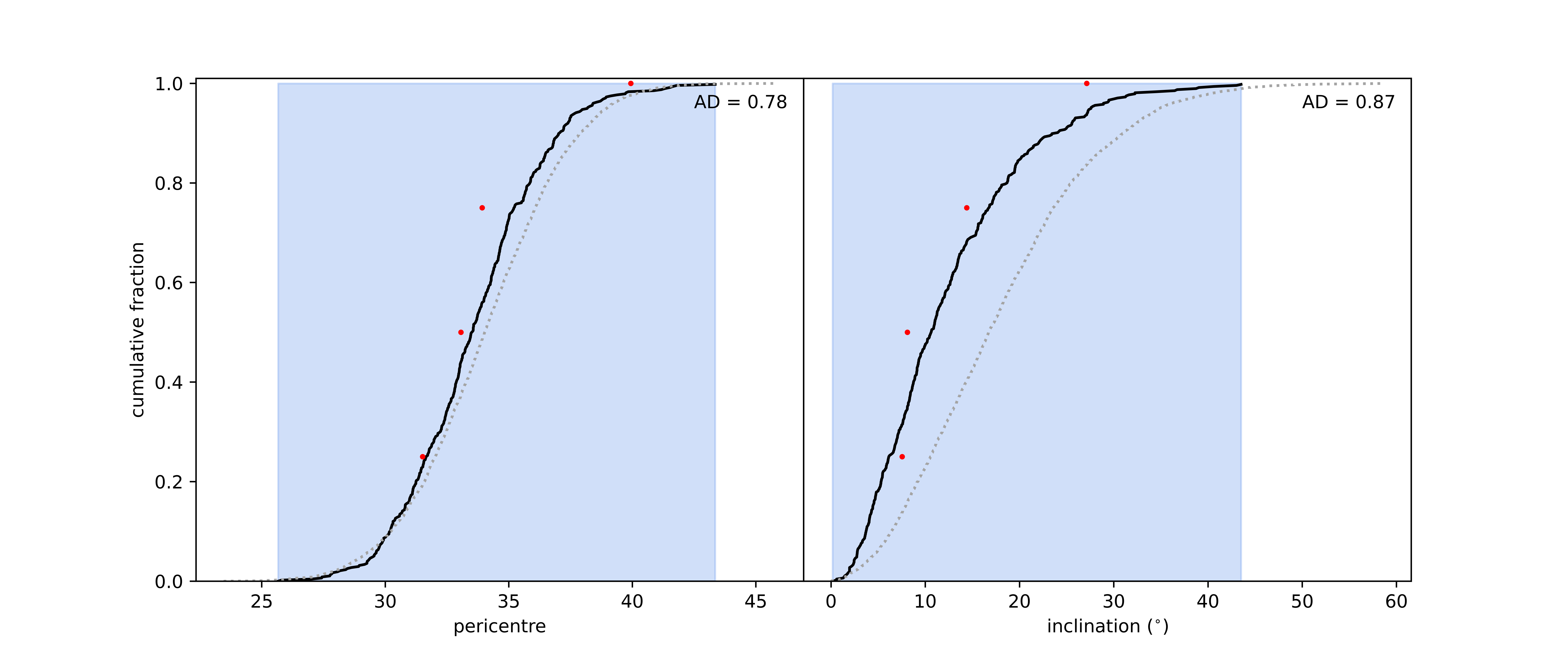}
    \caption{The best $n$:2 parameterized orbital model (see Table~\ref{tab:cumudist2}), shown as cumulative plots in pericenter distance (left) and inclination (right). Red markers indicate real OSSOS++ TNOs in $n$:2 resonances (excluding the 5:2 as discussed in the text), the grey dotted line is the intrinsic underlying model distribution, and the black line is the distribution of simulated detections. The blue box highlights the range of possible model values: the real TNOs must fall within this range for the model to be valid. These distributions are not rejectable with AD values of 0.78 and 0.87.}
    \label{fig:n2cumu}
\end{figure}

We had to find a balance between not over-fitting our small number of data points, while still realistically modelling similar but independent orbital distributions for different resonances.  
We decided to independently vary $q_c$, $q_w$, and $\sigma_i$, while all other orbital parameters were drawn from previously-fit distributions as discussed above.  
In order to have enough TNOs in each group for statistical analysis we grouped each set of $n$:$k$ resonances with the same $k$ together, except the 
$n$:1 resonances which were evaluated separately due to having more known TNOs. 
The parameterizations were evaluated by measuring cumulative distributions in batches of $n$:2, $n$:3, etc, and calculating the Anderson-Darling test statistic on the survey-biased parameterization vs.\ the real data (Figure~\ref{fig:n2cumu}). 

If scattering-sticking is the dominant pathway by which these distant resonances are populated \citep[e.g.][]{Yu:2018}, than we would expect the $q$ distributions to be similar across all of these resonances.  
While there are indications that this may be true for some of the distant resonances, to allow the best model with our limited data, we allow the distributions to be different for different groups of resonances. 

Due to the small number of detections, it was not difficult to find the part of parameter space that produced statistically acceptable models.  
We tested a number of different values for the width and center of the perihelion distribution (Equation~\ref{eq:q}) and the width of the inclination distribution (Equation~\ref{eq:i}) for each set of resonant populations.
We tested values of $\sigma_i$ ranging from 12$^\circ$ to 25$^\circ$, $q_c$ from 34AU to 39AU, and $q_w$ from 3AU to 4.5AU.
The final parameters and rejectability statistics for these parameterizations are given in Table \ref{tab:cumudist2}, where we have selected the least-rejectable parameters as determined via testing over these ranges. 
We stress that we cannot complete a comprehensive statistical fitting of our three parameters, which would be uninformative due to the small number of detections; we can say that these parameters are non-rejectable, and present a plausible orbital distribution for each resonance. 
With additional detections of TNOs in these resonances in the future, these parameterizations can be refined in a statistically robust manner.  
The exact values of $\sigma_i$, $q_c$, and $q_w$ do not greatly impact our population results compared to the already large uncertainties from the survey simulator analysis (see Table~\ref{tab:popsdist2}), which are dominated by the small numbers of detected objects.
For completeness, we also tested two other modelling strategies, and were unable to statistically reject either (see Appendix~\ref{sec:other}).  Though the $n$:1 resonances are each fitted individually, these values have little deviation from the values that are acceptable in the simplest model we tried (see Appendix~\ref{sec:other}), with $\sigma_i$ significantly larger than the $\sigma_i$'s which fit all other resonances. The increase in pericentre location $q_c$ in Table~\ref{tab:cumudist2} is an artifact of the objects orbiting at increasing semi-major axis, and is not otherwise significant. 

\begin{deluxetable}{cccccc}										
\tablecaption{Parameters for the Best Orbital Model.\label{tab:cumudist2}}										
\tablewidth{0pt}										
\tablehead{										
\colhead{Resonance}	&	\colhead{$\sigma_i$}	&	\colhead{AD$_i^a$}	&	\colhead{$q_c$}	&	\colhead{$q_w$}	&	\colhead{AD$_q$} \\
	&	\colhead{[$^{\circ}$]}	&		&	\colhead{[AU]}	&	\colhead{[AU]}	&	
}										
\startdata										
3:1	&	20	&	95\%	&	36	&	3	&	9\% \\
4:1	&	20	&	26\%	&	38	&	3	&	38\%\\
5:1	&	25	&	34\%	&	38	&	4	&	97\%\\
9:1	&	25	&	25\%	&	40	&	4	&	27\%\\ \hline
n:2	&	14.5	&	87\%	&	34	&	3	&	78\% \\
n:3	&	14.5	&	40\%	&	37.5	&	3	&	66\%  \\
n:4	&	25	&	25\%	&	37.5	&	3.5	&	36\% \\
n:5	&	20	&	46\%	&	38	&	4	&	70\% \\
n:6	&	14.5	&	97\%	&	37.5	&	3.5	&	68\% \\
n:8	&	14.5	&	87\%	&	37.5	&	3.5	&	98\% \\
\enddata										
\tablenotetext{a}{AD here refers to the rejectability of the model, and is based off the bootstrapped Anderson-Darling statistic. A value $\geq 5\%$ is considered \textit{not} statistically rejectable at the 95\% confidence level.}										
\end{deluxetable}

\subsection{Populations} \label{sec:pops}
 
Table~\ref{tab:popsdist2} gives the calculated populations in each of these resonances for our best orbital parameterizations, as determined by the rejectability statistic.  
In order to determine population estimates, a general method often used by the OSSOS team is employed \citep[e.g.][]{KavelaarsL3}. 
To obtain a population estimate with uncertainties, the Survey Simulator generated simulated objects from the parameterization above and determined whether each simulated object would have been detectable with OSSOS++, until the number of model detections matched the number of real TNO detections in each resonance. This process was repeated 1000 times with different random number seeds. The median number of objects generated within the Survey Simulator in order to obtain the desired number of simulated detections is taken to be the most likely population value; with 95\% confidence limits determined from the 25th and 975th highest values in the ordered list of the 1000 generated simulated populations.

Because our real detections include no $H_r$ magnitudes fainter than 10 (Appendix~\ref{sec:tnolist}), we run our population calculations down to a limiting magnitude of $H_r<10$.
We then scale these populations using the best-fit divot power-law $H$-distribution described in Section~\ref{sec:parameterized} to $H_r<8.66$, which corresponds to a diameter $D\gtrsim100$~km assuming an albedo of 0.04, and allows for easy comparison with previous population estimates from OSSOS++.
These are the populations given in Table~\ref{tab:popsdist2}.

A population estimate for the 5:2 resonance is also included in Table \ref{tab:popsdist2}. This population was determined by \citet{matthews_2019} through similar methods to those used in this work. 
We note that these best-fit populations agree with the other parameterizations tested in Appendix~\ref{sec:other} within error bars, highlighting that exact choice of parameterization has little effect on the population constraints.

\begin{deluxetable}{ccccccc}																	
\tablecaption{Populations for all distant resonances beyond the 2:1.\label{tab:popsdist2}}																	
\tablewidth{0pt}																	
\tablehead{																	
\colhead{Resonance}	&	\colhead{Semimajor}	&	\colhead{Number of}	&	\colhead{$q_c$}	&	\colhead{$q_w$}	&	\colhead{$\sigma_i$}	&	\colhead{Median Population} \\					
	&	\colhead{axis [AU]}	&	\colhead{detections}	&	\colhead{[AU]}	&	\colhead{[AU]}	&	\colhead{[$^{\circ}$]}	&	\colhead{($H_r<8.66$)}					
}																	
\startdata																	
3:1	&	62.5	&	12	&	36	&	3	&	20	&	17000	$^{+	11000	}_{-	8000	}$ \\
4:1	&	75.7	&	5	&	38	&	3	&	20	&	13000	$^{+	15000	}_{-	8000	}$ \\
5:1	&	87.9	&	3	&	38	&	4	&	25	&	11000	$^{+	19000	}_{-	8000	}$ \\
9:1	&	130.0	&	2	&	40	&	4	&	25	&	18000	$^{+	39000	}_{-	15000	}$ \\ \hline
5:2$^a$	&	55.3	&	29	&	39	&	5	&	17	&	6600	$^{+	4100	}_{-	3000	}$ \\
7:2	&	69.3	&	2	&	34	&	3	&	14.5	&	2300	$^{+	5400	}_{-	1900	}$ \\
9:2	&	81.9	&	1	&	34	&	3	&	14.5	&	1100	$^{+	6000	}_{-	1100	}$ \\
23:2	&	153.1	&	1	&	34	&	3	&	14.5	&	4000	$^{+	15000	}_{-	4000	}$ \\ \hline
7:3	&	52.9	&	1	&	37.5	&	3	&	14.5	&	3000	$^{+	5000	}_{-	2300	}$ \\
8:3	&	57.8	&	2	&	37.5	&	3	&	14.5	&	2300	$^{+	5000	}_{-	2000	}$ \\
10:3	&	67.1	&	1	&	37.5	&	3	&	14.5	&	1400	$^{+	6000	}_{-	1400	}$ \\ \hline
11:4	&	59.0	&	2	&	37.5	&	3.5	&	25	&	3900	$^{+	9000	}_{-	3400	}$ \\
15:4	&	72.5	&	2	&	37.5	&	3.5	&	25	&	2600	$^{+	12000	}_{-	2500	}$ \\
17:4	&	78.8	&	1	&	37.5	&	3.5	&	25	&	3100	$^{+	12000	}_{-	3000	}$ \\
27:4	&	107.3	&	1	&	37.5	&	3.5	&	25	&	5000	$^{+	23000	}_{-	4800	}$ \\ \hline
11:5	&	50.8	&	2	&	38	&	4	&	20	&	2100	$^{+	4900	}_{-	1800	}$ \\
12:5	&	53.9	&	2	&	38	&	4	&	20	&	2400	$^{+	5600	}_{-	2000	}$ \\
13:5	&	56.8	&	3	&	38	&	4	&	20	&	1200	$^{+	4800	}_{-	1200	}$ \\
24:5	&	85.5	&	1	&	38	&	4	&	20	&	2500	$^{+	11000	}_{-	2400	}$ \\ \hline
13:6	&	50.3	&	5	&	37.5	&	3.5	&	14.5	&	1700	$^{+	4300	}_{-	1400	}$ \\
23:6	&	73.6	&	1	&	37.5	&	3.5	&	14.5	&	1800	$^{+	7400	}_{-	1800	}$ \\ \hline
17:8	&	49.7	&	2	&	37.5	&	3.5	&	14.5	&	700	$^{+	3100	}_{-	700	}$ \\
35:8	&	80.4	&	1	&	37.5	&	3.5	&	14.5	&	2600	$^{+	11000	}_{-	2500	}$ \\ \hline \hline
TOTAL	&		&		&		&		&		&	110,000	$^{+	240,000	}_{-	82,000	}$ \\
\enddata																	
\tablenotetext{a}{The population estimate for the 5:2 was constrained by \citet{matthews_2019}. It is included here for completeness.}																	
\end{deluxetable}

 \section{Discussion} \label{sec:disc}

\subsection{Comparison to Previous Population Measurements}

The total population in all of these distant, high-order resonances is actually quite significant: more than 100,000 TNOs for $H_r>8.66$ (albeit with large uncertainty).  This is a factor of several larger than the population of the hot classical Kuiper Belt to a similar size limit \citep{Petit:2011,Kavelaars2021}, and an order of magnitude larger than the current scattering disk \citep{Lawler2018scattering}, though we note that at the 95\% lower limit this is on the same order as the current scattering disk. If distant resonances are dominantly occupied by scattering-sticking, our models may have a propensity to over-estimate the populations in these resonances, 
possibly due to using parameterizations based on the orbital distributions of the better-measured inner resonances. However, if we consider these large populations to be accurate to the order of magnitude, then this suggests that either a) the scattering-sticking mechanism is not the dominant mechanism by which bodies end up in resonance; or b) that the amount of bodies which get stuck in resonances must greatly increase with distance beyond Neptune, particularly in the $n$:1 resonances, as is suggested by the simulations in \citet{Yu:2018}. 

Our population measurements are compared with previous resonance population measurements in Table~\ref{tab:gladvscrom}.
The previous measurements are in agreement within the (rather large) error bars, but are in general lower.
The fact that these population values are higher than previously considered in large analyses like \citet{gladman2012resonant} may be attributed to a better expectation of large populations and inclination widths of distant TNOs. Much of this indeed comes from the fact that the inclination widths for the postulated distributions are no longer so tightly bounded as they were in \cite{gladman2012resonant}, where distant resonances were assumed to require a similar inclination width to the closer-in resonances.
The 5:1 and 9:1 resonances were modelled in more detail in \citet{Pike:2015} and \citet{Volk:2018}, respectively.  It is reassuring that our simpler parameterization is consistent with populations measured in these previous analyses that were focused on only one resonance.

\begin{deluxetable}{cccccc}										
\tablecaption{A comparison of our measured populations with measurements from the literature.\label{tab:gladvscrom}}										
\tablewidth{0pt}										
\tablehead{										
\colhead{Resonance}	&	\colhead{\citet{gladman2012resonant}$^a$}	&	\colhead{\cite{Lawler2013}$^a$}	&	\colhead{\cite{Pike:2015}}	&	\colhead{\cite{Volk:2018}}	&	This work
}										
\startdata										
3:1	&	$4000^{+9000}_{-3000}$	&	-	&	-	&	-	&	17,000$^{+11,000}_{-8000}$ \\
4:1	&	-	&	$<$16 000	&	-	&	-	&	13,000$^{+15,000}_{-8000}$ \\
5:1	&	$8000^{+34000}_{-7000}$	&	-	&	$23~000^{+40000}_{-17000}$	&	-	&	12 400$^{+12100}_{-8900}$\\
9:1	&	-	&	-	&	-	&	$11~000^{+19000}_{-7000}$	&	18 100$^{+28300}_{-12900}$\\
7:3	&	$4000^{+8000}_{-3000}$	&	-	&	-	&	-	&	3000$^{+6000}_{-2000}$\\
\enddata										
\tablenotetext{}{All populations are measured to $H_r<8.66$, and error bars and upper limits are 95\% confidence levels.}	
\tablenotetext{a}{Populations have been converted from $g$-band to $r$-band as described in Section~\ref{sec:smallnum}}
\label{tab:gladvscromp}
\end{deluxetable}

\subsection{Comparison with Published Migration and Scattering-Sticking Models}

Only a few Neptune migration models exist where the test particle orbits are publicly available, dynamically classified, and of sufficient quality to test against debiased observational data.  Here we look at the populations in two published studies: \citet{Pikeetal2017}\footnote{The dynamically classiﬁed model from \citet{Pikeetal2017} is publicly
available at \url{http://doi:10.11570/16.0009}.} which dynamically classified Nice Model-style simulations from \citet{Brasseretal2012}, and \citet{Lawler:2019}\footnote{The dynamically classiﬁed model from \citet{Lawler:2019} is publicly available at \url{http://doi:10.11570/19.0008}.} which dynamically classified a set of simulations using grainy (G) or smooth migration (Sm) over different timescales: fast 10~Myr (F) or slow 100~Myr (S) from \citet{KaibSheppard2016}.

In order to make a comparison between these models and our population measurements, we compare population ratios of objects in different resonances.  
We compare all the populations to the 8:3 resonance, since it is included in all the migration and scattering-sticking models, it is at moderate semimajor axis within the distant resonances discussed in this work, and it does not have an extreme population in any way.
To calculate our population ratios, we use the best populations for each resonance from Table~\ref{tab:popsdist2}.
We calculate the errors on these population ratios by randomly drawing from the distributions of simulated populations that are consistent with the number of real OSSOS++ detections (see Section~\ref{sec:pops}) and calculating the ratio 1000 times.  We then use this distribution of ratios to calculate 95\% confidence limits, which are given as error bars in Tables~\ref{tab:resrat} and \ref{tab:scat} and in Figure~\ref{fig:popratios}. 

In Table \ref{tab:resrat} and Figure~\ref{fig:popratios} we show the population ratios between resonances that we measured in this work, and resonance population ratios from the five Neptune migration models.  None of the five migration models provide populations ratios that are consistent with those seen in our models which are derived from real detections in the outer Solar System measured by OSSOS++: the migration models all severely underpopulate the 3:1 and 4:1 resonances relative to the higher-order 7:2 and 8:3 resonances. 

In addition to Neptune migration models, we also compare our population ratios to those calculated from scattering-sticking populations derived in \citet{Yu:2018}. 
The predicted population ratios match ours well (Table~\ref{tab:scat}), much better than the predictions from Neptune migration models (Figure~\ref{fig:popratios}).  We find the low-$k$ $j$:$k$ resonances are actually significantly more populated than higher-$k$, as predicted by scattering-sticking in \citet{Yu:2018}.  However, our absolute population estimates are significantly higher than those of \citet{Yu:2018}, which are scaled off the total current scattering population. This discrepancy may be due to our large populations being spurious, but even at 95\% lower limits our populations remain significantly higher than predicted by \citet{Yu:2018}. 

With precise orbits and the resulting population measurements from large surveys like OSSOS++, DES \citep{DES2020}, and soon LSST \citep{lsstbook}, cosmogonic models of Neptune migration and transneptunian orbital population and its subsequent evolution over 4.5 Gyr, including carefully accounting for scattering-sticking, must produce orbits and relative populations that are in agreement with this precise data and those populations derived from models which correct for survey biases. Ideally, cosmogonic models should produce publicly available test particle orbits that can then be tested against various survey data sets using the survey simulator, as we have done here for OSSOS.

\begin{deluxetable}{lcccc}													
\tablecaption{Population ratios to the 8:3\label{tab:resrat}}													
\tablewidth{0pt}													
\tablehead{													
\colhead{Model} & \colhead{8:3} & \colhead{7:2} & \colhead{3:1} & \colhead{4:1} 													
}													
\startdata													
This work	&	1.0	&	1.0	$^{+5.0}_{-0.8}$	&	7.9	$^{+23.5}_{-5.0}$	&	5.9	$^{+23.1}_{-4.3}$	\\	\hline
\citet{Lawler:2019} GF	&	1.0	&	0.7		&	1.9		&	0.4		\\	
\citet{Lawler:2019} GS	&	1.0	&	0.6		&	1.0		&	0.6		\\	
\citet{Lawler:2019} SmF	&	1.0	&	0.4		&	0.9		&	0.6		\\	
\citet{Lawler:2019} SmS	&	1.0	&	0.7		&	1.0		&	0.6		\\	
\citet{Pikeetal2017} Nice model	&	1.0	&	1.3		&	0.7		&	0.6		\\	\hline
\citet{Yu:2018} Scattering-sticking	&	1.0	&	2.3		&	3.6		&	5.4		\\	
\enddata
\tablenotetext{}{GF = Grainy Fast ($\sim$10~Myr) migration, GS = Grainy Slow ($\sim$100~Myr) migration, SmF = Smooth Fast migration, SmS = Smooth Slow migration.}	
\end{deluxetable}																												
\begin{deluxetable}{lcccc}										
\tablecaption{Scattering-sticking population comparison\label{tab:scat}}										
\tablewidth{0pt}										
\tablehead{										
 & \multicolumn{2}{c}{population ratio with 8:3} \\										
\colhead{$j$:$k$} & \colhead{this work} & \colhead{\citet{Yu:2018}}										
}										
\startdata										
3:1	&	7.4	$^{+	23.5	}_{-	5.0	}$	&	3.6	\\
4:1	&	5.7	$^{+	23.1	}_{-	4.3	}$	&	5.4	\\
5:1	&	4.8	$^{+	27.7	}_{-	3.5	}$	&	9.7	\\
7:2	&	1.0	$^{+	5.0	}_{-	0.8	}$	&	2.3	\\
9:2	&	0.5	$^{+	4.4	}_{-	0.5	}$	&	6.1	\\
7:3	&	1.3	$^{+	6.6	}_{-	1.1	}$	&	1.0	\\
8:3	&	1.0						&	1.0	\\
10:3	&	0.6	$^{+	4.3	}_{-	0.6	}$	&	1.9	\\
11:4	&	1.7	$^{+	8.8	}_{-	1.6	}$	&	0.4	\\
15:4	&	1.1	$^{+	8.5	}_{-	1.3	}$	&	2.3	\\
17:4	&	1.3	$^{+	8.3	}_{-	1.4	}$	&	2.4	\\
11:5	&	0.9	$^{+	4.6	}_{-	0.8	}$	&	0.4	\\
12:5	&	1.0	$^{+	5.9	}_{-	0.8	}$	&	0.4	\\
13:5	&	0.5	$^{+	4.0	}_{-	0.5	}$	&	0.4	\\
24:5	&	1.1	$^{+	9.4	}_{-	1.0	}$	&	2.1	\\
13:6	&	0.7	$^{+	3.6	}_{-	0.6	}$	&	0.2	\\
23:6	&	0.8	$^{+	11.3	}_{-	1.6	}$	&	0.7	\\
17:8	&	0.3	$^{+	2.7	}_{-	0.3	}$	&	0.1	\\
35:8	&	1.1	$^{+	7.7	}_{-	1.1	}$	&	0.4	\\
\enddata										
\end{deluxetable}	

\subsection{Scattering-sticking is important for distant resonances}  \label{sec:scat}
  
\begin{figure}
     
     \includegraphics[width=\linewidth]{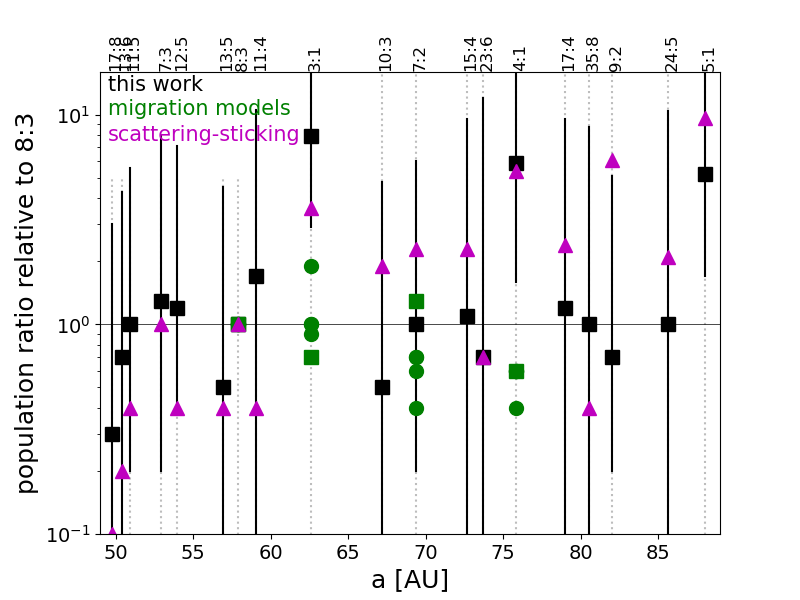}
     \caption{Measured and predicted populations, as a ratio to the 8:3 population (as in Tables~\ref{tab:resrat} and \ref{tab:scat}).  Black squares with error bars are our 95\% confidence limits on our measured population ratios from OSSOS++, green circles are population predictions from Neptune migration models dynamically classified in \citet{Lawler:2019}, green squares are population predictions from a Nice Model simulation dynamically classified in \citet{Pikeetal2017}, and magenta triangles are scattering-sticking population predictions from \citet{Yu:2018}.}
     \label{fig:popratios}
 \end{figure}

Besides resonance capture or sweeping during Neptune migration, TNOs can enter resonances at later times by temporarily sticking to them from the dynamically unstable scattering population.  
As evidenced in Table~\ref{tab:resrat}, it is difficult to populate the distant resonances with Neptune migration alone (at least, in those models which are publicly available and of sufficient resolution). \citet{Yu:2018} ran extensive simulations of a large scattering population for up to 1~Gyr, and calculated the population of scattering TNOs which should be stuck in different resonances at any moment.
These simulations provide the closest match to our measured resonant population ratios (Figure~\ref{fig:popratios} and Table~\ref{tab:scat}).

Because of the very large error bars on our population measurements, we can only say that the \citet{Yu:2018} population ratio predictions are closest to our measurements out of the models compared (Table~\ref{tab:resrat} and Figure~\ref{fig:popratios}), and not definitively rule out any of the models.  
\citet{Volk:2018} also provides a careful comparison between the \citet{Yu:2018} model and the $n$:1 resonance populations as measured by OSSOS++, and find that resonant-sticking populations should increase with semi-major axis, but this is the first time one has been able to compare across many different high-order resonances.
It is entirely possible that the Neptune migration models simply did not include enough scattering particles due to computational limits; higher resolution simulations that include both Neptune migration and large numbers of scattering particles may be able to reproduce our measurements even more closely in the future. Rather than requiring exact matches, we will follow the suggestions of \citet{Yu:2018} in proposing that the best model is the one which best recreates the populations of objects across the full disk. In our results, this is evaluated in the ratio comparisons across multiple resonances as in Table~\ref{tab:resrat}, with the best fit being to \citet{Yu:2018}.
 
Although these models are simple approximations of the distant transneptunian belt, they function as solid generalizations. The capacity to model the distant transneptunian belt is as of yet heavily stunted by the low number of discoveries and lack of precise orbital measurements, such that there are many arbitrary distributions which can recreate the one or two objects observed. The parameterizations presented here are just a starting point for comparison with realistic dynamical modelling to understand the past history of our solar system.  Further discoveries will provide the capability to model the Solar System to greater accuracy.

 \section{Conclusions}
 
This work is the first to measure the orbital distributions and populations of many of the high-order resonances within the distant transneptunian belt. The populations and parameters modelled here provide incremental improvements in testing and understanding of the distant small body populations in our solar system. The populations within these resonances are large and suggest a large fraction derived from scattering-sticking; thus this work serves as a push to create future theoretical migration models which more accurately match reality.
 
Because many of the higher order resonances only have 1-3 detections in them, we cannot statistically test them individually. As such, our best parameterization followed on the principle that all of these very distant objects arose from a similar mechanism and thus should have similar perihelion and inclination distributions. We assume distributions similar to those used to describe closer-in resonances, which may have very different orbital element distributions if TNOs are dominantly emplaced in these resonances by a different process than in the distant resonances. We also assume no interaction from any additional unknown giant planet whose presence, if real, would change the dynamics of the distant resonances \citep[e.g.,][]{Malhotra:P9}. Our results are provided in absolute populations with 95\% confidence limits, and are compared to previous results via population ratios.
 
The populations we measure in the distant resonances are generally larger than originally anticipated, both as compared with Neptune migration models, and compared to previous observational constraints on their populations (though they match observational measurements within the 95\% confidence limits). 
The \citet{Yu:2018} scattering-sticking model does the best job of reproducing the population ratios we measured, but more discoveries are needed to determine if scattering-sticking alone can create these large populations and if the true orbital distributions match scattering-sticking predictions, or if a primordial, swept-up/captured population is also needed.
These large populations in the distant resonances are a challenge that must be reproduced in the next generation of theoretical solar system formation and migration models.
A hopeful primary source of these new discoveries is the Vera C.~Rubin Observatory, which is predicted to discover thousands of new TNOs, and with careful orbital measurements, many will be found to reside in distant orbital resonances. 
We are hopeful that despite the growing threat of megaconstellations of artificial satellites, enough TNOs will be discovered and tracked to be able to create more precise models, which will provide further understanding of the history of our solar system.

\begin{acknowledgements}
The authors acknowledge the sacred nature of Maunakea and appreciate the opportunity to observe from the mountain.  BLC and SML acknowledge and appreciate conducting our research on Canadian Treaty 4 land, the territories of the n\^{e}hiyawak, Anih\v{s}in\={a}p\={e}k, Dakota, Lakota, and Nakoda, and the homeland of the M\'{e}tis/Michif Nation.

The authors wish to thank the two anonymous referees for providing helpful comments and suggestions to improve and clarify this paper.

SML and BG acknowledge the support of the Natural Sciences and Engineering Research Council of Canada (NSERC), Discovery Grants RGPIN-2020-04111 and RGPIN-2018-04895, respectively. 
KV acknowledges support from NASA (grants 80NSSC19K0785 and 80NSSC21K0376) and NSF (grant AST-1824869).
MTB appreciates support by the Rutherford Discovery Fellowships from New Zealand Government funding, administered by the Royal Society Te Ap\={a}rangi.
This work was supported by the Programme National de Plantologie (PNP) of CNRS-INSU co-funded by CNES (JMP).
This research used the Canadian Advanced Network For Astronomy Research (CANFAR) operated in partnership by the  Canadian Astronomy Data Centre and The Digital Research Alliance of Canada with support from the National Research Council of Canada the Canadian Space Agency, CANARIE and the Canadian Foundation for Innovation.
\end{acknowledgements}

\facilities{CFHT (MegaPrime), CANFAR} 
\software{matplotlib \citep{Hunter2007}, 
scipy \citep{Jonesetal2001},
Mercury \citep{Chambers2001}, 
SWIFT \citep{LevisonDuncan1994},
OSSOS Survey Simulator \citep{Lawler2018survey,Petit:2018}, 
Numpy \citep{numpy}, 
Python \citep{PYTHON} 
}

\appendix
\section{Additional simple parameterizations} \label{sec:other}

As the high-order resonances had few detected TNOs in each, we tested three methods in order to create parameterized orbital distributions.  One of these three was presented as our best-fit model in the main text (see Table~\ref{tab:popsdist2}), and for completeness, we also include our other two models in this Appendix.

\underline{Unified Distribution Model}: We take the entire set of $n$:$k$ detections, with $k>1$, and attempt to model them with the same values of $\sigma_i$, $q_c$, and $q_w$. $n$:1's were modelled separately because the inclination widths which were acceptable for all other resonances did not pass the rejectability test with the $n$:1's.
We allow these parameters to vary in $\sigma_i$ (12$^\circ$ to 17$^\circ$), $q_c$ (34AU to 39AU), and $q_w$ (3AU to 4.5AU) and find the values that produce non-rejectable simulated detections for all resonances simultaneously. This more conservative range in $\sigma_i$ is used as it most accurately fits the unified distribution of all measured TNOs; higher inclinations as used in our current model for $n$:4 and $n$:5 were not within the reasonable range for this distribution, and would lead to spurious estimates.

The best parameters and rejectability (AD) test values are summarised in Table~\ref{tab:cumudist}. Note that while this was not our preferred model, this parameterization is statistically non-rejectable.
The populations measured with this parameterization were generally slightly larger than those given in Table~\ref{tab:popsdist2}, but are consistent within the 95\% uncertainties given.

\underline{Minimized Population Model}:  We also model the resonances with three or greater detections with assumptions that allow a minimum population estimate. We fix the perihelion distance distribution to just include the real observed objects and find the narrowest $\sigma_i$ that results in simulated detected objects that include the observed range of inclinations.
The populations measured with this parameterization are (as expected) lower than those presented in Table~\ref{tab:cumudist2}, but still within the 95\% uncertainties.
While this model is also simple and statistically non-rejectable, it is not our preferred model, as we have no reason to believe that the inclination distributions should be minimized for any dynamical reason.

\begin{figure}
    \centering
    \includegraphics[width=\textwidth]{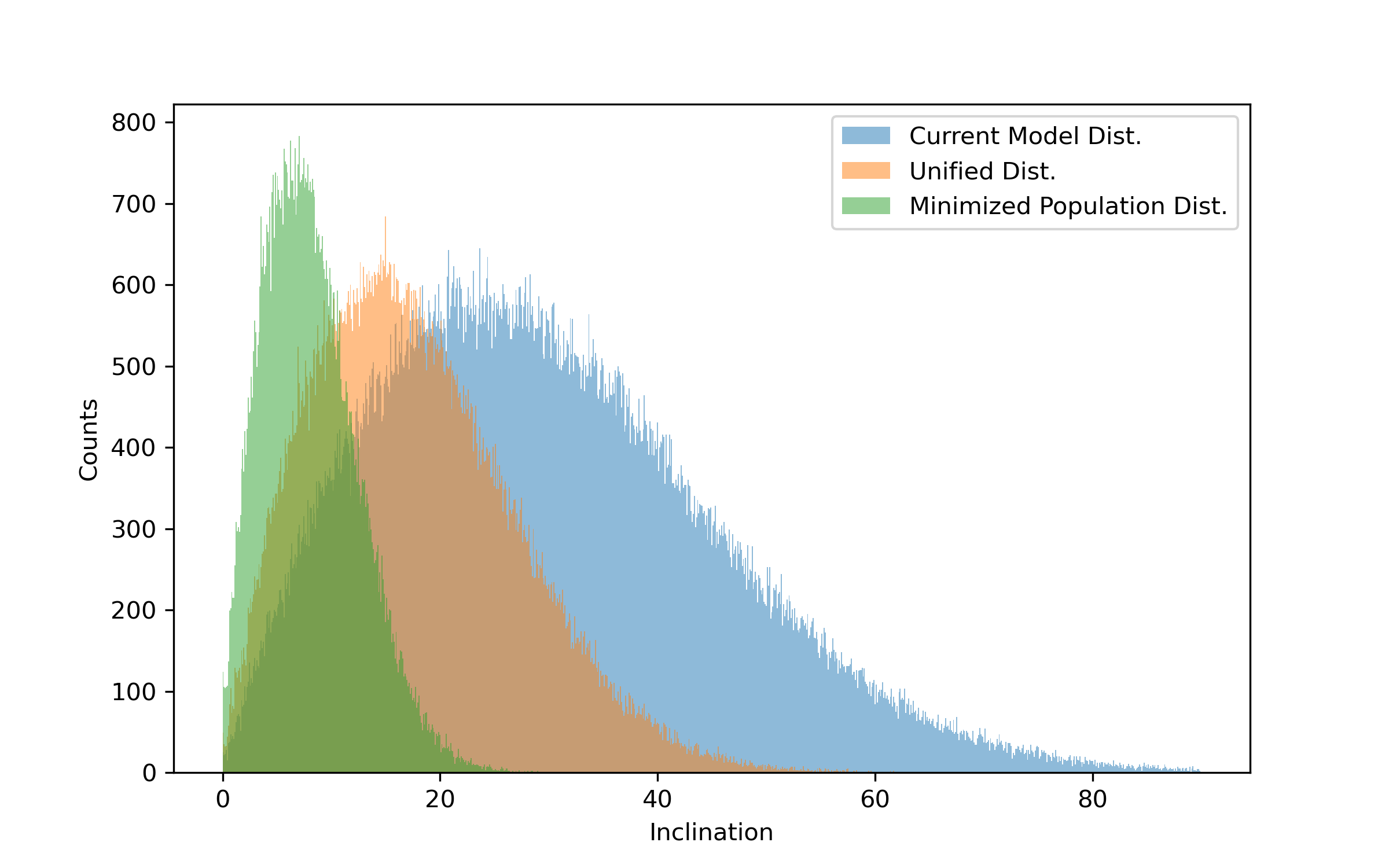}
\caption{All three models are compared through a histogram detailing the variance of their inclination in the 11:4 resonance. Although in each case the distribution is determined by sin(i) times a Gaussian (Equation \ref{eq:i}), the width of that Gaussian differs in each case as per the given model. The current model (blue) has a population of N=3900, Unified Distribution (peach) has a population of N=2800, and Minimized Population Distribution (green) has a population of N=1400. 
}
    \label{fig:inchist3models}
\end{figure}

The three models are compared for one example resonance in Figure~\ref{fig:inchist3models}. 

\begin{deluxetable}{ccccccc}										
\tablecaption{Simplest paramaterization: the Unified Distribution Model.\label{tab:cumudist}}										
\tablewidth{0pt}										
\tablehead{										
\colhead{Resonance}	&	\colhead{$\sigma_i$}	&	\colhead{AD$_i$}	&	\colhead{$q_c$}	&	\colhead{$q_w$}	&	\colhead{AD$_q$} \\
	&	\colhead{[$^{\circ}$]}	&		&	\colhead{[AU]}	&	\colhead{[AU]}	&	
}										
\startdata										
3:1	&	25	&	15\%	&	37.35	&	3.35	&	15\%\\
4:1	&	25	&	34\%	&	37.35	&	3.35	&	31\%\\
5:1	&	25	&	29\%	&	37.35	&	3.35	&	97\%\\
9:1	&	25	&	28\%	&	37.35	&	3.35	&	6\%\\ \hline
n:1	&	25	&	61\%	&	37.35	&	3.35	&	70\% \\ \hline
n:2	&	14.5	&	87\%	&	37.35	&	3.35	&	19\% \\
n:3	&	14.5	&	34\%	&	37.35	&	3.35	&	47\% \\
n:4	&	14.5	&	7\%	&	37.35	&	3.35	&	33\% \\
n:5	&	14.5	&	18\%	&	37.35	&	3.35	&	65\% \\
n:6	&	14.5	&	94\%	&	37.35	&	3.35	&	59\% \\
n:8	&	14.5	&	79\%	&	37.35	&	3.35	&	85\% \\\hline
n:k	&	14.5	&	35\%	&	37.35	&	3.35	&	65\%\\
\enddata										
\end{deluxetable}

\section{Distant Resonant TNOs Detected by OSSOS++} \label{sec:tnolist}

Table~\ref{tab:ossoslist} lists measured orbital elements and $H_r$ magnitudes for the distant resonant TNOs detected by OSSOS++ and used in our analysis. 

\begin{deluxetable}{cccccccccc}																			
\tablecaption{Distant resonant TNOs detected by OSSOS++  \label{tab:ossoslist}}																			
\tablehead{ 																			
\colhead{MPC}&  \colhead{Survey}&    \colhead{Res}&   \colhead{a}&    \colhead{e}&  \colhead{i}&   \colhead{band$^a$} & \colhead{Apparent}&   \colhead{Absolute}& \colhead{Secure$^b$} \\																			
\colhead{designation}&  &    &   \colhead{[AU]}&    &  \colhead{[$^{\circ}$]}&    &   \colhead{mag.} & \colhead{$H$-mag}& 																			
}																			
\startdata																			
K13U17B	&	OSSOS	&	3:1	&	62.507291	&	0.286532	&	32.857	&	r	&	23.71	&	6.78	&	S	\\
K15VG6M	&	OSSOS	&	3:1	&	63.022881	&	0.441183	&	0.813	&	r	&	24.06	&	8.19	&	S	\\
K15VG6N	&	OSSOS	&	3:1	&	62.717552	&	0.480977	&	6.03	&	r	&	24.94	&	8.97	&	S	\\
K15KH3Y	&	OSSOS	&	3:1	&	62.646269	&	0.435875	&	8.997	&	r	&	24.68	&	8.56	&	S	\\
K15G55A	&	OSSOS	&	3:1	&	62.632298	&	0.387251	&	17.958	&	r	&	24.28	&	8.19	&	S	\\
K15G55B	&	OSSOS	&	3:1	&	62.065678	&	0.410198	&	10.783	&	r	&	24.73	&	8.29	&	I	\\
K15RR8A	&	OSSOS	&	3:1	&	62.61022	&	0.427743	&	11.344	&	r	&	25.06	&	8.8	&	S	\\
K15RR7Z	&	OSSOS	&	3:1	&	62.239671	&	0.444722	&	5.827	&	r	&	24.94	&	6.58	&	I	\\
K04VD0U	&	CFEPS	&	3:1	&	62.193747	&	0.428055	&	8.024	&	g	&	23.92	&	6.95	&	S	\\
K11Uf1S	&	Alexandersen	&	3:1	&	62.430542	&	0.499834	&	22.04	&	r	&	24.03	&	9.1	&	S	\\
K11Uf1R	&	Alexandersen	&	3:1	&	61.841493	&	0.433397	&	26.582	&	r	&	24	&	8.32	&	I	\\
K11Uf1Q	&	Alexandersen	&	3:1	&	62.44225	&	0.404531	&	40.4	&	r	&	23.9	&	6.75	&	I	\\ \hline
K15VG6O	&	OSSOS	&	4:1	&	75.698691	&	0.491114	&	11.273	&	r	&	24.81	&	5.64	&	I	\\
K15VG6P	&	OSSOS	&	4:1	&	75.207299	&	0.524308	&	21.293	&	r	&	24.63	&	8.3	&	I	\\
K15KH3Z	&	OSSOS	&	4:1	&	75.625945	&	0.591465	&	20.878	&	r	&	24.74	&	9.49	&	S	\\
K15RR8B	&	OSSOS	&	4:1	&	75.654926	&	0.438846	&	27.694	&	r	&	24.68	&	7.66	&	I	\\
K11Uf1P	&	Alexandersen	&	4:1	&	75.785556	&	0.491166	&	13.435	&	r	&	24.2	&	8.34	&	S	\\ \hline
K03YH9Q	&	CFEPS	&	5:1	&	88.411905	&	0.578658	&	20.874	&	g	&	23.34	&	7.33	&	I	\\
K07F51N	&	HiLat	&	5:1	&	87.493815	&	0.618771	&	23.237	&	r	&	23.2	&	7.17	&	I	\\
K07L38F	&	HiLat	&	5:1	&	87.569525	&	0.555223	&	35.825	&	r	&	22.53	&	5.54	&	I	\\ \hline
K07Th4C	&	OSSOS	&	9:1	&	129.944666	&	0.695216	&	26.468	&	r	&	23.21	&	7.13	&	S	\\
K15KH2E	&	OSSOS	&	9:1	&	129.800344	&	0.66001	&	38.361	&	r	&	24.67	&	8.2	&	S	\\ \hline
K13GD6Y	&	OSSOS	&	5:2	&	55.536335	&	0.414113	&	10.877	&	r	&	22.94	&	7.32	&	S	\\
K13GD6S	&	OSSOS	&	5:2	&	55.625089	&	0.385455	&	6.978	&	r	&	23.92	&	8.34	&	S	\\
K13U17D	&	OSSOS	&	5:2	&	55.462608	&	0.395744	&	26.372	&	r	&	24.26	&	8.86	&	S	\\
K13U17O	&	OSSOS	&	5:2	&	55.005779	&	0.388514	&	3.833	&	r	&	23.71	&	8.21	&	S	\\
K13J64F	&	OSSOS	&	5:2	&	55.427838	&	0.449733	&	8.785	&	r	&	23.86	&	8.97	&	S	\\
K13J64K	&	OSSOS	&	5:2	&	55.246946	&	0.408298	&	11.077	&	r	&	22.94	&	7.69	&	S	\\
K14UM9S	&	OSSOS	&	5:2	&	55.260678	&	0.398085	&	3.902	&	r	&	23.18	&	7.95	&	S	\\
K14UN1O	&	OSSOS	&	5:2	&	55.413839	&	0.237022	&	23.156	&	r	&	24.63	&	7.89	&	S	\\
K14UN0A	&	OSSOS	&	5:2	&	55.333326	&	0.231957	&	12.705	&	r	&	24.11	&	6.4	&	S	\\
K15VG7E	&	OSSOS	&	5:2	&	55.286713	&	0.386226	&	29.716	&	r	&	24.53	&	9.18	&	S	\\
K15VG7D	&	OSSOS	&	5:2	&	55.470402	&	0.414859	&	16.275	&	r	&	24.56	&	8.88	&	S	\\
K15VG7F	&	OSSOS	&	5:2	&	55.50662	&	0.272825	&	36.662	&	r	&	24.31	&	8.19	&	S	\\
K15KH4F	&	OSSOS	&	5:2	&	55.635637	&	0.469353	&	8.976	&	r	&	24.79	&	10	&	S	\\
K15KH4G	&	OSSOS	&	5:2	&	55.474877	&	0.456026	&	18.611	&	r	&	23.61	&	8.43	&	S	\\
K15KH4D	&	OSSOS	&	5:2	&	55.318614	&	0.420513	&	17.013	&	r	&	23.84	&	8.63	&	S	\\
K15KH4H	&	OSSOS	&	5:2	&	55.343603	&	0.428349	&	10.336	&	r	&	23.2	&	7.64	&	S	\\
K15KH4E	&	OSSOS	&	5:2	&	55.437365	&	0.289947	&	24.065	&	r	&	23.49	&	7.47	&	S	\\
K15KH4C	&	OSSOS	&	5:2	&	55.551475	&	0.407396	&	8.407	&	r	&	24.03	&	7.74	&	S	\\
K15G55M	&	OSSOS	&	5:2	&	55.423937	&	0.436625	&	9.031	&	r	&	24.46	&	9.45	&	S	\\
K15G55N	&	OSSOS	&	5:2	&	55.263876	&	0.412585	&	8.325	&	r	&	24.42	&	5.84	&	S	\\
K15RR8H	&	OSSOS	&	5:2	&	55.483721	&	0.385306	&	5.038	&	r	&	24.97	&	6.89	&	I	\\
K00F08E	&	CFEPS	&	5:2	&	55.286289	&	0.402014	&	5.869	&	g	&	22.56	&	6.97	&	S	\\
K04E96G	&	CFEPS	&	5:2	&	55.55032	&	0.422907	&	16.213	&	g	&	23.5	&	8.36	&	S	\\
\enddata																			
\tablenotetext{a}{CFEPS primarily used $g$-band, while other OSSOS++ surveys used $r$-band. Apparent and $H$ magnitudes are given in the band for each survey as listed.}																			
\tablenotetext{b}{S = Securely in resonance, I = Insecure, diagnosed from 10~Myr integrations of 3 clones within semimajor axis orbital uncertainty.}																			
\end{deluxetable}

\begin{deluxetable}{cccccccccc}																			
\tablecaption{CONTINUED: Distant resonant TNOs detected by OSSOS++  \label{tab:ossoslist2}}																			
\tablehead{ 																			
\colhead{MPC}&  \colhead{Survey}&    \colhead{Res}&   \colhead{a}&    \colhead{e}&  \colhead{i}&   \colhead{band$^a$} & \colhead{Apparent}&   \colhead{Absolute}& \colhead{Secure$^b$} \\																			
\colhead{designation}&  &    &   \colhead{[AU]}&    &  \colhead{[$^{\circ}$]}&    &   \colhead{mag.} & \colhead{$H$-mag}& 																			
}																			
\startdata	
K02G32P	&	CFEPS	&	5:2	&	55.386888	&	0.421953	&	1.559	&	g	&	22.14	&	7.02	&	S	\\
K04H79O	&	CFEPS	&	5:2	&	55.205527	&	0.411661	&	5.624	&	g	&	23.59	&	7.82	&	S	\\
K04K18Z	&	CFEPS	&	5:2	&	55.418575	&	0.381906	&	22.645	&	g	&	24.05	&	8.64	&	S	\\
K07L38G	&	HiLat	&	5:2	&	55.452064	&	0.434019	&	32.579	&	r	&	22.93	&	7.68	&	S	\\
K12UH7J	&	Alexandersen	&	5:2	&	55.196603	&	0.433338	&	15.632	&	r	&	24.1	&	8.62	&	I	\\
K11Uf1T	&	Alexandersen	&	5:2	&	55.668479	&	0.406186	&	6.42	&	r	&	24.53	&	7.45	&	I	\\ \hline
K15VG7R	&	OSSOS	&	7:2	&	69.462751	&	0.54637	&	14.394	&	r	&	24.75	&	8.19	&	S	\\
K15KH4O	&	OSSOS	&	7:2	&	69.532871	&	0.524499	&	8.103	&	r	&	23.12	&	7.84	&	S	\\ \hline
K15RO5R	&	OSSOS	&	9:2	&	81.728544	&	0.584918	&	7.552	&	r	&	21.76	&	3.6	&	I	\\ \hline
K15KG3H	&	OSSOS	&	23:2	&	153.081338	&	0.739108	&	27.137	&	r	&	24.73	&	7.57	&	I	\\ \hline
K13J64N	&	OSSOS	&	7:3	&	53.051438	&	0.287664	&	7.74	&	r	&	23.84	&	7.96	&	S	\\
K01XP4T	&	CFEPS	&	7:3	&	52.921424	&	0.322046	&	0.518	&	g	&	23.48	&	7.77	&	S	\\
K02CO8Z	&	CFEPS	&	7:3	&	53.039438	&	0.38913	&	5.466	&	g	&	24.08	&	8.5	&	S	\\ \hline
K14UM8E	&	OSSOS	&	8:3	&	57.935212	&	0.402251	&	8.798	&	r	&	24.23	&	8.25	&	S	\\
K14UM8K	&	OSSOS	&	8:3	&	58.053135	&	0.353779	&	7.683	&	r	&	23.66	&	7.23	&	I	\\ \hline
K11Uf2Q	&	Alexandersen	&	10:3	&	67.321317	&	0.482854	&	16.712	&	r	&	23.03	&	7.5	&	I	\\ \hline
K13J64H	&	OSSOS	&	11:4	&	59.196664	&	0.383985	&	13.73	&	r	&	22.7	&	5.6	&	I	\\
K15VG7O	&	OSSOS	&	11:4	&	58.908495	&	0.406897	&	17.629	&	r	&	24.03	&	7.98	&	I	\\ \hline
K14UM9V	&	OSSOS	&	15:4	&	72.39879	&	0.498662	&	27.694	&	r	&	24.25	&	8.44	&	I	\\ \hline
K15RR8L	&	OSSOS	&	17:4	&	79.061455	&	0.534497	&	17.629	&	r	&	24.37	&	8.53	&	S	\\ \hline
K04PB2B	&	OSSOS	&	27:4	&	107.515853	&	0.67128	&	15.434	&	r	&	22.99	&	7.39	&	I	\\ \hline
K15RR8U	&	OSSOS	&	11:5	&	50.847654	&	0.246819	&	27.196	&	r	&	23.26	&	6.55	&	I	\\
K04Q29H	&	CFEPS	&	11:5	&	50.859008	&	0.229217	&	12.01	&	g	&	23.54	&	7.54	&	I	\\ \hline
K15RR8S	&	OSSOS	&	12:5	&	53.895011	&	0.411296	&	13.803	&	r	&	23.88	&	8.67	&	S	\\
K02CM4Y	&	CFEPS	&	12:5	&	53.891995	&	0.34651	&	15.733	&	g	&	22.3	&	6.7	&	S	\\ \hline
K14UM8B	&	OSSOS	&	13:5	&	56.921255	&	0.329063	&	7.434	&	r	&	23.71	&	7.9	&	S	\\ \hline
K15VG7Q	&	OSSOS	&	24:5	&	85.581312	&	0.556146	&	28.694	&	r	&	25.06	&	8.15	&	I	\\ \hline
K15VG7T	&	OSSOS	&	13:6	&	50.372549	&	0.190425	&	6.207	&	r	&	24.71	&	8.56	&	S	\\
K15RR8V	&	OSSOS	&	13:6	&	50.375358	&	0.277451	&	9.376	&	r	&	24.28	&	8.34	&	S	\\ \hline
K12UH7R	&	Alexandersen	&	23:6	&	73.78739	&	0.491581	&	16.353	&	r	&	24.23	&	7.73	&	I	\\ \hline
K14UN0B	&	OSSOS	&	17:8	&	49.675999	&	0.276738	&	11.489	&	r	&	23.89	&	8.24	&	I	\\ \hline
K15VG7S	&	OSSOS	&	35:8	&	80.484191	&	0.514092	&	7.111	&	r	&	24.89	&	8.94	&	S	\\
\enddata																			
\tablenotetext{a}{CFEPS primarily used $g$-band, while other OSSOS++ surveys used $r$-band. Apparent and $H$ magnitudes are given in the band for each survey as listed.}																			
\tablenotetext{b}{S = Securely in resonance, I = Insecure, diagnosed from 10~Myr integrations of 3 clones within semimajor axis orbital uncertainty.}																			
\end{deluxetable}


\begin{thebibliography}{}
\expandafter\ifx\csname natexlab\endcsname\relax\def\natexlab#1{#1}\fi
\providecommand{\url}[1]{\href{#1}{#1}}
\providecommand{\dodoi}[1]{doi:~\href{http://doi.org/#1}{\nolinkurl{#1}}}
\providecommand{\doeprint}[1]{\href{http://ascl.net/#1}{\nolinkurl{http://ascl.net/#1}}}
\providecommand{\doarXiv}[1]{\href{https://arxiv.org/abs/#1}{\nolinkurl{https://arxiv.org/abs/#1}}}

\bibitem[{Alexandersen {et~al.}(2016)Alexandersen, Gladman, Kavelaars, Petit,
  Gwyn, Shankman, \& Pike}]{Alexandersen:2016}
Alexandersen, M., Gladman, B., Kavelaars, J., {et~al.} 2016, The Astronomical
  Journal, 152, 111

\bibitem[{Anderson \& Darling(1954)}]{Anderson-Darling}
Anderson, T.~W., \& Darling, D.~A. 1954, Journal of the American Statistical
  Association, 49, 765, \dodoi{10.1080/01621459.1954.10501232}
  
\bibitem[Bannister et al.(2016)]{Bannister:2016} Bannister, M.~T., Alexandersen, M., Benecchi, S.~D., et al.\ 2016, \aj, 152, 212. \dodoi{10.3847/0004-6256/152/6/212}

\bibitem[{Bannister {et~al.}(2018)Bannister, Gladman, Kavelaars, Petit, Volk,
  Chen, Alexandersen, Gwyn, Schwamb, Ashton, {et~al.}}]{Bannister:2018}
Bannister, M.~T., Gladman, B.~J., Kavelaars, J., {et~al.} 2018, The
  Astrophysical Journal Supplement Series, 236, 18

\bibitem[{{Bernardinelli} {et~al.}(2020){Bernardinelli}, {Bernstein}, {Sako},
  {Liu}, {Saunders}, {Khain}, {Lin}, {Gerdes}, {Brout}, {Adams}, {Belyakov},
  {Somasundaram}, {Sharma}, {Locke}, {Franson}, {Becker}, {Napier},
  {Markwardt}, {Annis}, {Abbott}, {Avila}, {Brooks}, {Burke}, {Carnero Rosell},
  {Carrasco Kind}, {Castander}, {da Costa}, {De Vicente}, {Desai}, {Diehl},
  {Doel}, {Everett}, {Flaugher}, {Garc{\'\i}a-Bellido}, {Gruen}, {Gruendl},
  {Gschwend}, {Gutierrez}, {Hollowood}, {James}, {Johnson}, {Johnson},
  {Krause}, {Kuropatkin}, {Maia}, {March}, {Miquel}, {Paz-Chinch{\'o}n},
  {Plazas}, {Romer}, {Rykoff}, {S{\'a}nchez}, {Sanchez}, {Scarpine}, {Serrano},
  {Sevilla-Noarbe}, {Smith}, {Sobreira}, {Suchyta}, {Swanson}, {Tarle},
  {Walker}, {Wester}, {Zhang}, \& {DES Collaboration}}]{DES2020}
{Bernardinelli}, P.~H., {Bernstein}, G.~M., {Sako}, M., {et~al.} 2020, \apjs,
  247, 32, \dodoi{10.3847/1538-4365/ab6bd8}

\bibitem[{{Brasser} {et~al.}(2012){Brasser}, {Schwamb}, {Lykawka}, \&
  {Gomes}}]{Brasseretal2012}
{Brasser}, R., {Schwamb}, M.~E., {Lykawka}, P.~S., \& {Gomes}, R.~S. 2012,
  \mnras, 420, 3396, \dodoi{10.1111/j.1365-2966.2011.20264.x}

\bibitem[{Brown(2001)}]{Brown2001}
Brown, M.~E. 2001, The Astronomical Journal, 121, 2804

\bibitem[{{Chambers}(2001)}]{Chambers2001}
{Chambers}, J.~E. 2001, \icarus, 152, 205, \dodoi{10.1006/icar.2001.6639}

\bibitem[{Chen {et~al.}(2019)Chen, Gladman, Volk, Murray-Clay, Lehner,
  Kavelaars, Wang, Lin, Lykawka, Alexandersen, {et~al.}}]{Chen:2019}
Chen, Y.-T., Gladman, B., Volk, K., {et~al.} 2019, The Astronomical Journal,
  158, 214

\bibitem[{Collaboration {et~al.}(2009)Collaboration, Abell, Allison, Anderson,
  Andrew, Angel, Armus, Arnett, Asztalos, Axelrod, Bailey, Ballantyne, Bankert,
  Barkhouse, Barr, Barrientos, Barth, Bartlett, Becker, Becla, Beers,
  Bernstein, Biswas, Blanton, Bloom, Bochanski, Boeshaar, Borne, Bradac,
  Brandt, Bridge, Brown, Brunner, Bullock, Burgasser, Burge, Burke, Cargile,
  Chandrasekharan, Chartas, Chesley, Chu, Cinabro, Claire, Claver, Clowe,
  Connolly, Cook, Cooke, Cooray, Covey, Culliton, de~Jong, de~Vries,
  Debattista, Delgado, Dell'Antonio, Dhital, Stefano, Dickinson, Dilday,
  Djorgovski, Dobler, Donalek, Dubois-Felsmann, Durech, Eliasdottir, Eracleous,
  Eyer, Falco, Fan, Fassnacht, Ferguson, Fernandez, Fields, Finkbeiner,
  Figueroa, Fox, Francke, Frank, Frieman, Fromenteau, Furqan, Galaz, Gal-Yam,
  Garnavich, Gawiser, Geary, Gee, Gibson, Gilmore, Grace, Green, Gressler,
  Grillmair, Habib, Haggerty, Hamuy, Harris, Hawley, Heavens, Hebb, Henry,
  Hileman, Hilton, Hoadley, Holberg, Holman, Howell, Infante, Ivezic, Jacoby,
  Jain, R, Jedicke, Jee, Jernigan, Jha, Johnston, Jones, Juric, Kaasalainen,
  Styliani, Kafka, Kahn, Kaib, Kalirai, Kantor, Kasliwal, Keeton, Kessler,
  Knezevic, Kowalski, Krabbendam, Krughoff, Kulkarni, Kuhlman, Lacy, Lepine,
  Liang, Lien, Lira, Long, Lorenz, Lotz, Lupton, Lutz, Macri, Mahabal,
  Mandelbaum, Marshall, May, McGehee, Meadows, Meert, Milani, Miller, Miller,
  Mills, Minniti, Monet, Mukadam, Nakar, Neill, Newman, Nikolaev, Nordby,
  O'Connor, Oguri, Oliver, Olivier, Olsen, Olsen, Olszewski, Oluseyi, Padilla,
  Parker, Pepper, Peterson, Petry, Pinto, Pizagno, Popescu, Prsa, Radcka,
  Raddick, Rasmussen, Rau, Rho, Rhoads, Richards, Ridgway, Robertson, Roskar,
  Saha, Sarajedini, Scannapieco, Schalk, Schindler, Schmidt, Schmidt,
  Schneider, Schumacher, Scranton, Sebag, Seppala, Shemmer, Simon, Sivertz,
  Smith, Smith, Smith, Spitz, Stanford, Stassun, Strader, Strauss, Stubbs,
  Sweeney, Szalay, Szkody, Takada, Thorman, Trilling, Trimble, Tyson, Berg,
  Berk, VanderPlas, Verde, Vrsnak, Walkowicz, Wandelt, Wang, Wang, Warner,
  Wechsler, West, Wiecha, Williams, Willman, Wittman, Wolff, Wood-Vasey,
  Wozniak, Young, Zentner, \& Zhan}]{lsstbook}
Collaboration, L.~S., Abell, P.~A., Allison, J., {et~al.} 2009, LSST Science
  Book, Version 2.0.
\newblock \doarXiv{0912.0201}

\bibitem[{{Gladman} {et~al.}(2008){Gladman}, {Marsden}, \&
  {Vanlaerhoven}}]{Gladman2008}
{Gladman}, B., {Marsden}, B.~G., \& {Vanlaerhoven}, C. 2008, {Nomenclature in
  the Outer Solar System}, ed. M.~A. {Barucci}, H.~{Boehnhardt}, D.~P.
  {Cruikshank}, A.~{Morbidelli}, \& R.~{Dotson}, 43

\bibitem[{Gladman {et~al.}(2012)Gladman, Lawler, Petit, Kavelaars, Jones,
  Parker, Van~Laerhoven, Nicholson, Rousselot, Bieryla,
  {et~al.}}]{gladman2012resonant}
Gladman, B., Lawler, S., Petit, J.-M., {et~al.} 2012, The Astronomical Journal,
  144, 23

\bibitem[{Hahn \& Malhotra(2005)}]{Hahn:2005}
Hahn, J.~M., \& Malhotra, R. 2005, The Astronomical Journal, 130, 2392

\bibitem[{Hunter(2007)}]{Hunter2007}
Hunter, J.~D. 2007, Computing In Science \& Engineering, 9, 90

\bibitem[{{Jones} {et~al.}(2001){Jones}, {Oliphant}, {Peterson}, \&
  Others}]{Jonesetal2001}
{Jones}, E., {Oliphant}, T., {Peterson}, P., \& Others. 2001, SciPy: Open
  source scientific tools for Python.
\newblock \url{http://www.scipy.org/}

\bibitem[{Kaib {et~al.}(2011)Kaib, Quinn, {et~al.}}]{Kaib:2011}
Kaib, N.~A., Quinn, T., {et~al.} 2011, Icarus, 215, 491

\bibitem[{{Kaib} \& {Sheppard}(2016)}]{KaibSheppard2016}
{Kaib}, N.~A., \& {Sheppard}, S.~S. 2016, \aj, 152, 133,
  \dodoi{10.3847/0004-6256/152/5/133}

\bibitem[{{Kavelaars} {et~al.}(2021){Kavelaars}, {Petit}, {Gladman},
  {Bannister}, {Alexandersen}, {Chen}, {Gwyn}, \& {Volk}}]{Kavelaars2021}
{Kavelaars}, J., {Petit}, J.-M., {Gladman}, B., {et~al.} 2021, arXiv e-prints,
  arXiv:2107.06120.
\newblock \doarXiv{2107.06120}

\bibitem[{{Kavelaars} {et~al.}(2009){Kavelaars}, {Jones}, {Gladman}, {Petit},
  {Parker}, {Van Laerhoven}, {Nicholson}, {Rousselot}, {Scholl}, {Mousis},
  {Marsden}, {Benavidez}, {Bieryla}, {Campo Bagatin}, {Doressoundiram},
  {Margot}, {Murray}, \& {Veillet}}]{KavelaarsL3}
{Kavelaars}, J.~J., {Jones}, R.~L., {Gladman}, B.~J., {et~al.} 2009, \aj, 137,
  4917, \dodoi{10.1088/0004-6256/137/6/4917}

\bibitem[{Lawler(2013)}]{Lawler2013}
Lawler, S. 2013, PhD thesis, University of British Columbia, Vancouver, BC V6T
  1Z4, Canada

\bibitem[{Lawler \& Gladman(2013)}]{Lawler:2013}
Lawler, S., \& Gladman, B. 2013, The Astronomical Journal, 146, 6

\bibitem[{Lawler {et~al.}(2018{\natexlab{a}})Lawler, Shankman, Kavelaars,
  Alexandersen, Bannister, Chen, Gladman, Fraser, Gwyn, Kaib,
  {et~al.}}]{Lawler2018scattering}
Lawler, S., Shankman, C., Kavelaars, J., {et~al.} 2018{\natexlab{a}}, The
  Astronomical Journal, 155, 197

\bibitem[{Lawler {et~al.}(2019)Lawler, Pike, Kaib, Alexandersen, Bannister,
  Chen, Gladman, Gwyn, Kavelaars, Petit, {et~al.}}]{Lawler:2019}
Lawler, S., Pike, R., Kaib, N., {et~al.} 2019, The Astronomical Journal, 157,
  253

\bibitem[{Lawler {et~al.}(2018{\natexlab{b}})Lawler, Kavelaars, Alexandersen,
  Bannister, Gladman, Petit, \& Shankman}]{Lawler2018survey}
Lawler, S.~M., Kavelaars, J., Alexandersen, M., {et~al.} 2018{\natexlab{b}},
  Frontiers in Astronomy and Space Sciences, 5, 14

\bibitem[{{Levison} \& {Duncan}(1994)}]{LevisonDuncan1994}
{Levison}, H.~F., \& {Duncan}, M.~J. 1994, \icarus, 108, 18,
  \dodoi{10.1006/icar.1994.1039}

\bibitem[{Levison {et~al.}(2008)Levison, Morbidelli, VanLaerhoven, Gomes, \&
  Tsiganis}]{Levison:2008}
Levison, H.~F., Morbidelli, A., VanLaerhoven, C., Gomes, R., \& Tsiganis, K.
  2008, Icarus, 196, 258

\bibitem[{{Lin} {et~al.}(2021){Lin}, {Chen}, {Volk}, {Gladman}, {Murray-Clay},
  {Alexandersen}, {Bannister}, {Lawler}, {Ip}, {Lykawka}, {Kavelaars}, {Gwyn},
  \& {Petit}}]{Lin:2021}
{Lin}, H.~W., {Chen}, Y.-T., {Volk}, K., {et~al.} 2021, \icarus, 361, 114391,
  \dodoi{10.1016/j.icarus.2021.114391}

\bibitem[{{Lykawka} \& {Mukai}(2007)}]{LykawkaMukai2007}
{Lykawka}, P.~S., \& {Mukai}, T. 2007, \icarus, 189, 213,
  \dodoi{10.1016/j.icarus.2007.01.001}
  
\bibitem[Malhotra et al.(2016)]{Malhotra:P9} Malhotra, R., Volk, K., \& Wang, X.\ 2016, \apjl, 824, L22. doi:10.3847/2041-8205/824/2/L22

\bibitem[{Matthews(2019)}]{matthews_2019}
Matthews, A. 2019, A Dynamically Eroded Model of the 5:2 Trans-Neptunian Mean
  Motion Resonance

\bibitem[{{Morbidelli}(1997)}]{Morbidelli1997}
{Morbidelli}, A. 1997, \icarus, 127, 1, \dodoi{10.1006/icar.1997.5681}

\bibitem[Murray \& Dermott(2000)]{Murray} Murray, C.~D. \& Dermott, S.~F.\ 2000, ``Solar System Dynamics'', by C.D. Murray and S.F. Dermott. ISBN 0521575974. \url{http://www.cambridge.org/us/catalogue/catalogue.asp?isbn=0521575974}. Cambridge, UK: Cambridge University Press, 1999.
\dodoi{10.1017/CBO9781139174817}

\bibitem[Nesvorn{\'y} \& Roig(2001)]{Nesvorny:2001} Nesvorn{\'y}, D. \& Roig, F.\ 2001, \icarus, 150, 104. \dodoi{10.1006/icar.2000.6568}

\bibitem[{Nesvorn{\`y} \& Vokrouhlick{\`y}(2016)}]{Nesvorny:2016}
Nesvorn{\`y}, D., \& Vokrouhlick{\`y}, D. 2016, The Astrophysical Journal, 825,
  94

\bibitem[{Petit {et~al.}(2018)Petit, Kavelaars, Gladman, \&
  Alexandersen}]{Petit:2018}
Petit, J.-M., Kavelaars, J., Gladman, B., \& Alexandersen, M. 2018, Astrophysics Source Code Library, ascl-1805

\bibitem[{Petit {et~al.}(2011)Petit, Kavelaars, Gladman, Jones, Parker,
  Van~Laerhoven, Nicholson, Mars, Rousselot, Mousis, {et~al.}}]{Petit:2011}
Petit, J.-M., Kavelaars, J.~J., Gladman, B.~J., {et~al.} 2011, The Astronomical
  Journal, 142, 131

\bibitem[{Petit {et~al.}(2017)Petit, Kavelaars, Gladman, Jones, Parker,
  Bieryla, Van~Laerhoven, Pike, Nicholson, Ashby, {et~al.}}]{Petit:2017}
Petit, J.-M., Kavelaars, J., Gladman, B., {et~al.} 2017, The Astronomical
  Journal, 153, 236

\bibitem[{Pike {et~al.}(2015)Pike, Kavelaars, Petit, Gladman, Alexandersen,
  Volk, \& Shankman}]{Pike:2015}
Pike, R., Kavelaars, J., Petit, J.-M., {et~al.} 2015, The Astronomical Journal,
  149, 202

\bibitem[{{Pike} {et~al.}(2017){Pike}, {Lawler}, {Brasser}, {Shankman},
  {Alexandersen}, \& {Kavelaars}}]{Pikeetal2017}
{Pike}, R.~E., {Lawler}, S., {Brasser}, R., {et~al.} 2017, \aj, 153, 127,
  \dodoi{10.3847/1538-3881/aa5be9}
  
\bibitem[Pike \& Lawler(2017)]{PikeLawler:2017} Pike, R.~E. \& Lawler, S.~M.\ 2017, \aj, 154, 171. doi:10.3847/1538-3881/aa8b65

\bibitem[{{Shankman} {et~al.}(2013){Shankman}, {Gladman}, {Kaib}, {Kavelaars},
  \& {Petit}}]{Shankmanetal2013}
{Shankman}, C., {Gladman}, B.~J., {Kaib}, N., {Kavelaars}, J.~J., \& {Petit},
  J.~M. 2013, \apjl, 764, L2, \dodoi{10.1088/2041-8205/764/1/L2}

\bibitem[{Shankman {et~al.}(2016)Shankman, Kavelaars, Gladman, Alexandersen,
  Kaib, Petit, Bannister, Chen, Gwyn, Jakubik, {et~al.}}]{Shankman:2016}
Shankman, C., Kavelaars, J., Gladman, B., {et~al.} 2016, The Astronomical
  Journal, 151, 31

\bibitem[{{Tegler} {et~al.}(2016){Tegler}, {Romanishin}, {Consolmagno}, \&
  {J.}}]{Tegleretal2016}
{Tegler}, S.~C., {Romanishin}, W., {Consolmagno}, G.~J., \& {J.}, S. 2016, \aj,
  152, 210, \dodoi{10.3847/0004-6256/152/6/210}

\bibitem[{{Tiscareno} \& {Malhotra}(2009)}]{Tiscareno2009}
{Tiscareno}, M.~S., \& {Malhotra}, R. 2009, \aj, 138, 827,
  \dodoi{10.1088/0004-6256/138/3/827}

\bibitem[{Van Der~Walt {et~al.}(2011)Van Der~Walt, Colbert, \&
  Varoquaux}]{numpy}
Van Der~Walt, S., Colbert, S.~C., \& Varoquaux, G. 2011, Computing in science
  \& engineering, 13, 22

\bibitem[{van Rossum(1995)}]{PYTHON}
van Rossum, G. 1995, Python tutorial, Tech. Rep. CS-R9526, Centrum voor
  Wiskunde en Informatica (CWI), Amsterdam

\bibitem[Volk \& Malhotra(2013)]{Volk:2013} Volk, K. \& Malhotra, R.\ 2013, \dps

\bibitem[{Volk {et~al.}(2016)Volk, Murray-Clay, Gladman, Lawler, Bannister,
  Kavelaars, Petit, Gwyn, Alexandersen, Chen, {et~al.}}]{Volk:2016}
Volk, K., Murray-Clay, R., Gladman, B., {et~al.} 2016, The Astronomical
  Journal, 152, 23

\bibitem[{Volk {et~al.}(2018)Volk, Murray-Clay, Gladman, Lawler, Yu,
  Alexandersen, Bannister, Chen, Dawson, Greenstreet, {et~al.}}]{Volk:2018}
Volk, K., Murray-Clay, R.~A., Gladman, B.~J., {et~al.} 2018, The Astronomical
  Journal, 155, 260

\bibitem[{Yu {et~al.}(2018)Yu, Murray-Clay, \& Volk}]{Yu:2018}
Yu, T. Y.~M., Murray-Clay, R., \& Volk, K. 2018, The Astronomical Journal, 156,
  33

\end{thebibliography}
\end{document}